\documentclass[]{aa_mine}

\usepackage{graphicx}
\usepackage{txfonts}

\graphicspath{{imagens/}}

\newlength{\picwd}
\newlength{\thinpic}
\usepackage[english]{babel}
\usepackage[utf8]{inputenc}
\usepackage{natbib}
\usepackage{datetime}
\usepackage{xcolor}

\usepackage{textcomp}
\usepackage{gensymb}

\newcommand{\un}[1]{\ensuremath{\ \mathrm{#1}}}

\newcommand{\rsun}{\ensuremath{\ R_\odot}}
\begin{document}

\title{Flux-tube geometry and solar wind speed during an activity cycle}

\author{
  R. F. Pinto \inst{1,2}
  \and
  A. S. Brun \inst{3}
  \and
  A. P. Rouillard \inst{1,2}
}

\institute{
  Université de Toulouse; UPS-OMP; IRAP;  Toulouse, France
  \and
  CNRS; IRAP; 9 Av. colonel Roche, BP 44346, F-31028 Toulouse cedex 4, France\\
  \email{rui.pinto@irap.omp.eu}
   \and
   Laboratoire AIM Paris-Saclay, CEA/Irfu Universit\'e Paris-Diderot CNRS/ 
   INSU, 91191 Gif-sur-Yvette, France
}

\date{(\emph{manuscript version:} \today; \currenttime)}

 
  \abstract
  {
    The solar wind speed at $1\un{AU}$ shows cyclic variations in latitude and in time which reflect the evolution of the global background magnetic field during the activity cycle.
    It is commonly accepted that the terminal (asymptotic) wind speed in a given magnetic flux-tube is generally anti-correlated with its total expansion ratio, which motivated the definition of widely-used semi-empirical scaling laws relating one to the other.
    In practice, such scaling laws require \emph{ad-hoc} corrections (especially for the slow wind in the vicinities of streamer/coronal hole boundaries) and empirical fits to \emph{in-situ} spacecraft data.
    A predictive law based solely on physical principles is still missing.
  }
  %
  {
    We test whether the flux-tube expansion is the controlling factor of the wind speed at all phases of the cycle and at all latitudes (close and faraway from streamer boundaries) using a very large sample of wind-carrying open magnetic flux-tubes.
    We furthermore search for additional physical parameters based on the geometry of the coronal magnetic field which have an influence on the terminal wind flow speed.
  }
  %
  {
    We use numerical MHD simulations of the corona and wind coupled to a dynamo model to determine the properties of the coronal magnetic field and of the wind velocity (as a function of time and latitude) during a whole 11-year activity cycle.
   These simulations provide a large statistical ensemble of open flux-tubes which we analyse conjointly in order to identify relations of dependence between the wind speed and geometrical parameters of the flux-tubes which are valid globally (for all latitudes and moments of the cycle).
  }
  %
  {
    Our study confirms that the terminal (asymptotic) speed of the solar wind depends very strongly on the geometry of the open magnetic flux-tubes through which it flows.
    The total flux-tube expansion is more clearly anti-correlated with the wind speed for fast rather than for slow wind flows, and effectively controls the locations of these flows during solar minima.
    Overall, the actual asymptotic wind speeds attained -- specially those of the slow wind -- are also strongly dependent on field-line inclination and magnetic field amplitude at the foot-points.
    We suggest ways of including these parameters on future predictive scaling-laws for the solar wind speed.
  }
  %
  {}
  

  \maketitle

%
\section{Introduction}
\label{sec:introduction}


\begin{figure*}[!ht]
  \centering

  \includegraphics[height=0.77\picwd,clip,trim=0   0 335 0]{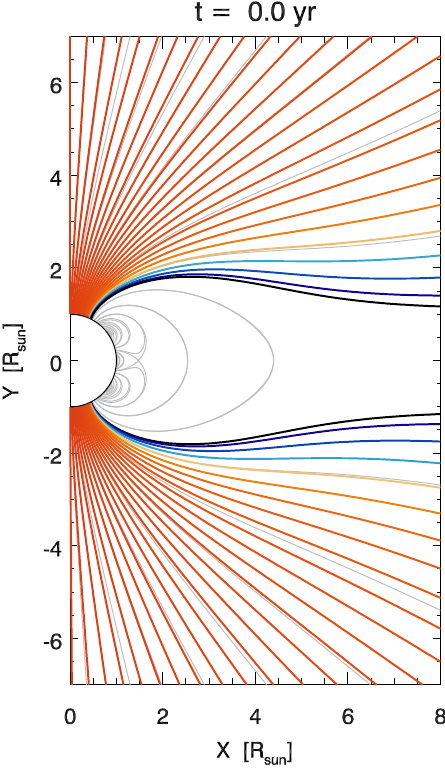}
  \includegraphics[height=0.77\picwd,clip,trim=18 0 335 0]{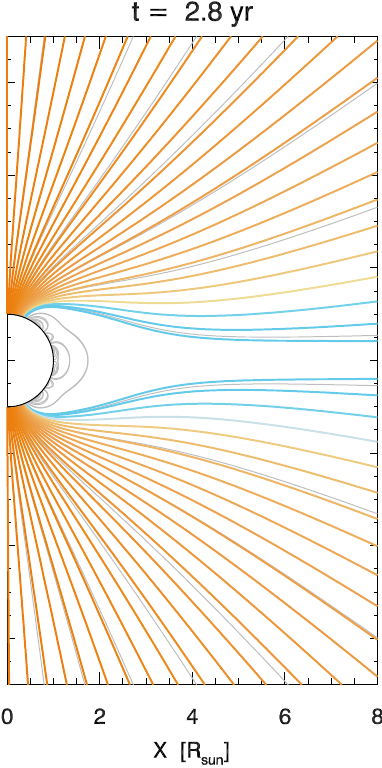}
  \includegraphics[height=0.77\picwd,clip,trim=18 0 335 0]{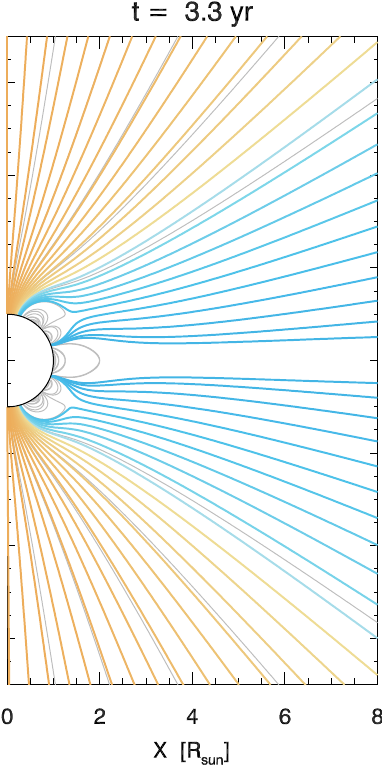}
  \includegraphics[height=0.77\picwd,clip,trim=18 0 335 0]{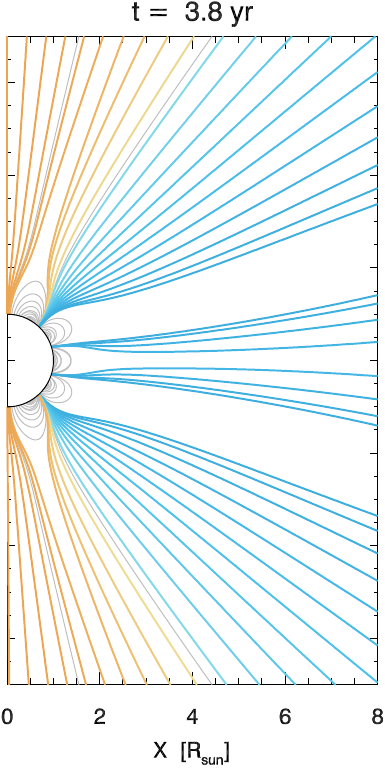}
  \includegraphics[height=0.77\picwd,clip,trim=18 0 335 0]{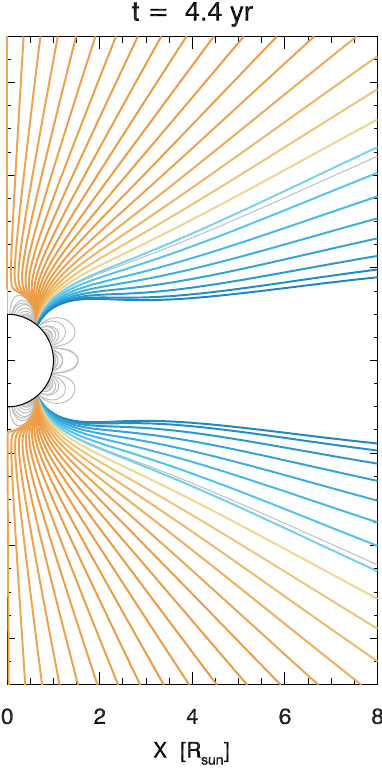}   
  \caption{
    Structure of the coronal magnetic field at five representative instants of the modelled solar cycle.
    The times are, from left to right, $t=0,\ 2.75\ , 3.30,\ 3.80,\ 4.40 \un{years}$ corresponding, respectively, to the activity minimum, the early rising phase, the late rising phase, the maximum, and the declining phase of the cycle.
    The field-lines are coloured according to the terminal speed of the wind (at the outer boundary of the domain, $r=15\un{\rsun}$) flowing along each one of them (with dark red and orange representing fast wind flows, and blue representing slow wind flows; \emph{cf.} colourbar in Fig. \ref{fig:intro-model-speed}).
    The grey lines represent closed field-lines, for completion.
  }
  \label{fig:intro-model-snaps}
\end{figure*}

The solar wind flow exhibits a large-scale distribution of fast and slow wind streams which evolves markedly during the solar activity cycle.
During solar minima, slow wind flows are essentially confined to a region between $20\degree$ and $-20\degree$ around the ecliptic plane and emanate from the vicinity of the streamer -- coronal hole boundaries (S/CH), while fast wind flows stream out from the polar coronal holes and fill all the polar and mid-latitude regions faraway from the Sun.
During solar maxima, these two wind components get mixed in latitude as a consequence of streamers (and pseudo-streamers) appearing at high solar latitudes and of coronal holes making incursions into the low-latitude regions \citep{mccomas_three-dimensional_2003}.
It is clear that the spatial (latitudinal) distribution of slow and fast wind flows follows closely the cyclic variations of the underlying coronal magnetic field structure, itself a consequence of the $11$ year cycle of the solar dynamo \citep{mccomas_weaker_2008,smith_solar_2011,richardson_solar_2008}.
This fact, together with the notion that the properties of the surface motions (assumed as energy sources for the heating and acceleration of the wind) are much more uniform across the solar disk than the amplitude of the wind flows above, suggests that the coronal environment causes for the segregation between fast and slow solar wind flows.
In particular, the wind terminal speeds seem to be determined to a great extent by the geometrical properties of the magnetic flux-tubes through which the solar wind flows \citep{wang_solar_1990}.
Theories that predict the solar wind speed often make use of simple parameters describing the variations of the cross-sections of the flux-tubes as a function of height, namely the expansion factor
\begin{equation}
  \label{eq:expans_definition_general}
  f = \frac{A_1}{A_0} \left(\frac{r_0}{r_1}\right)^2,
\end{equation}
where $A_0$ and $A_1$ are the cross-section of a given elemental flux-tube respectively at the surface of the Sun ($r=r_0$) and at some point higher in the corona ($r = r_1 > r_0$) above which the flux-tubes expand radially outwards (and not super-radially).
A radially expanding flux-tube has a total expansion ratio $f=1$, while a very strongly expanding flux-tube has $f >>1$.
The use of potential field extrapolations with source-surface (PFSS) from magnetogram data lead to associating $r_1$ with the radius of the source-surface, commonly placed at a fixed height of $r_{SS} = 2.5\un{\rsun}$, and to setting the expansion factor $f\equiv f_{SS}$ in respect to this height \citep{wang_solar_1990}.
This value for the source-surface radius was determined to be the one which produced the best match between the geometry of the extrapolated magnetic fields and the shapes of the coronal structures observed in white-light during solar eclipses, especially the size of the streamers and coronal hole boundaries \citep{wang_formation_2010,wang_coronal_2009}.
Matching quantities such as the open magnetic flux requires, however, defining $r_{SS}$ as a function of the solar activity \citep{lee_coronal_2011,arden_breathing_2014}, or more generally as a function of the properties of the global coronal magnetic field \citep{reville_solar_2015}.

\citet{suzuki_forecasting_2006} also suggested that the terminal wind speed would be better predicted by a combination of the expansion factor and the magnetic field amplitude at the foot-point of any given flux-tube or, equivalently, to the open magnetic flux.
Other authors also invoke empirically derived parameters such as the angular distance from the foot-point of a given magnetic flux-tube to the nearest streamer / coronal-hole (S/CH) boundary \citep[parameter $\theta_b$; ][]{arge_improved_2003,arge_stream_2004,mcgregor_distribution_2011}.
Recent studies by \citet{li_solar_2011} and \citet{lionello_application_2014} also brought forward that field-line curvature may have an impact on the wind speed based on solar wind simulations constrained by idealised magnetic geometries.
Analysis of Ulysses data combined with PFSS extrapolations by \citet{peleikis_investigation_2016} also suggests that field-line bending correlates to some extent with terminal wind speed.
Here, we analyse in deeper detail the distributions of terminal wind speed in respect to the geometrical properties of a very large set of individual flux-tubes. 
We use numerical method described by \citet{pinto_coupling_2011} to couple a solar dynamo to the corona and the solar wind in order to have access to a large sample of solar wind profiles representative of all solar latitudes and moments of the activity cycle.

\section{Methods}
\label{sec:methods}

\begin{figure}[]
  \centering
  \includegraphics[width=\picwd]{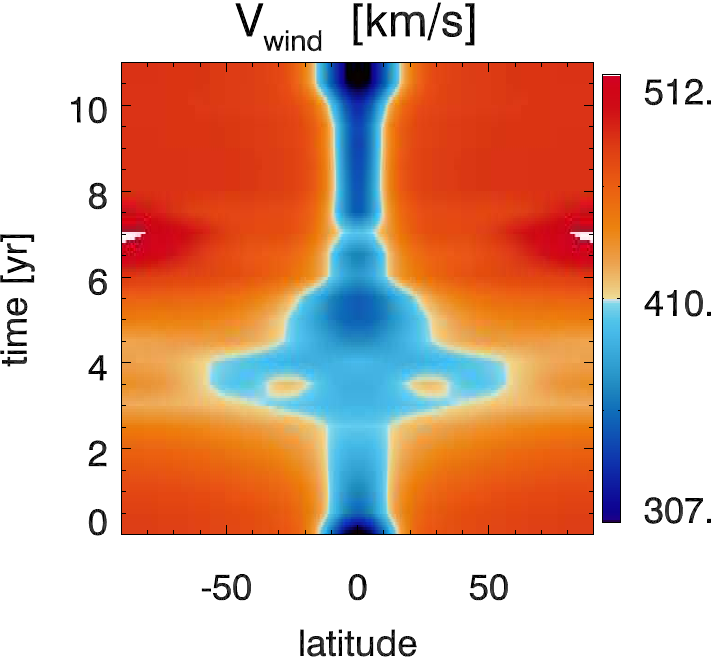}
  \caption{
    Time-latitude diagram of the terminal wind speed covering 11 years of the solar cycle (y-axis) and 180 degrees of latitude (from north to south pole, x-axis).
    The colour-scheme is the same as in the field-lines plotted in Fig. \ref{fig:intro-model-snaps}.
  }
  \label{fig:intro-model-speed}
\end{figure}

We use global-scale MHD simulations of the solar dynamo, corona and wind to investigate how the solar wind speed relates with flux-tube geometry at all latitudes and moments of the solar cycle.
The model provides us with maps of the time-varying coronal magnetic field and of the wind velocity in the meridional plane from the solar surface up to $15\rsun$ during an $11\un{yr}$ activity cycle.
The simulation method, described in detail in \citet{pinto_coupling_2011}, couples two 2.5D axisymmetric MHD codes.
The first one ---  STELEM \citep{jouve_role_2007} --- computes the temporal evolution of the surface magnetic field using a mean-field kinematic dynamo (with a Babcock-Leighton source term at the surface) driven by the meridional circulation and differential rotation in the convection zone.
The second code --- DIP \citep{grappin_alfven_2000} --- computes the temporal evolution of an MHD solar corona with a self-consistent wind.
The model assumes the corona and the wind to be isothermal, with a uniform coronal temperature $T_0 = 1.3 \un{MK}$ and a specific heat ratio $\gamma = 1$ (hence, we do not model the heating and cooling mechanisms in detail).
Figure \ref{fig:intro-model-snaps} represents some snapshots of the evolution of the coronal magnetic field during the solar cycle in our model.
The open magnetic field-lines are coloured according to the terminal speed of the wind flow which streams along each one of them, with dark red corresponding to fastest wind flows ($\sim 520\un{km/s}$ at $15\un{\rsun}$) and dark blue to the slowest wind flows ($\sim 250\un{km/s}$ at $15\un{\rsun}$).
The grey-lines are closed field-lines, with no bulk flow along them.
The first instant represented in the figure (on the panel to the left) corresponds to the minimum of activity, and shows a large equatorial streamer surrounded by two wide polar coronal holes.
A single equatorial current sheet extends from the top of the large streamer outwards.
Fast wind flows well inside the polar coronal holes, hence originating at high latitudes at the surface of the Sun and spreading out to lower latitudes higher up in the corona.
Slow wind flows are restricted to the regions nears the boundaries between the equatorial streamer and the polar coronal holes.
This general picture is maintained during the rising phase of the cycle (second and third panels of Fig. \ref{fig:intro-model-snaps}, up to about $3.5\un{yr}$), albeit with the slow wind spreading progressively into higher latitudes.
The fourth instant represented on Fig. \ref{fig:intro-model-snaps} corresponds roughly to the maximum of activity, just before the global polarity reversal.
At this stage, the large equatorial streamer has given place to multiple smaller streamers lying at low latitudes, and new high-latitude streamers have appeared.
Several current sheets are now present at different latitudes.
Slow wind flows now occupy a much wider range of latitudes, and the overall contrast between the maximum and minimum wind speeds became smaller.
A very moderate slow to fast wind gradient from the equator to the poles still persists, nevertheless.
The solar wind now originates at multiple and non-contiguous locations over the surface, but each with a latitudinal extent which is much smaller than those of the polar coronal holes present at the minimum of activity.
Other important topological features such as pseudo-streamers (different from streamers in that they are formed in unipolar regions in coronal holes and are not associated to an heliospheric current sheet) also appear both during the cycle rise and decay phases, as shown respectively in the third and fifth panels of the figure \citep[see also][]{pinto_coupling_2011}.

Fig. \ref{fig:intro-model-speed} shows the temporal evolution of the latitudinal distribution of fast/slow wind distributions at $15\un{\rsun}$ during the cycle in a time-latitude diagram.
The colour-scheme is the same as in Fig. \ref{fig:intro-model-snaps}, with orange tones representing fast wind flows and blue tones representing slow wind flows.
The diagram shows that the fast wind is more prevalent during activity minima, in opposition to the slow wind.
The width of the boundaries between slow and fast wind flows varies during the cycle.
The transition from slow to fast wind is much sharper during activity minima.
This fast wind - slow wind pattern shows good qualitative agreement with the IPS radio maps of \citet{manoharan_solar_2009} and
\citet{tokumaru_solar_2010}, and with the estimations by \citet{wang_sources_2006} using ULYSSES data and semi-empirical methods.
Overall, the contrast between the highest and the lowest wind speeds is highest at the minimum of activity and lowest at the maximum.
The wind speeds we obtain for the fast wind flows are, however, lower than those measured in the solar wind.
Adding additional sources of acceleration known to be efficient in the fast wind regime, such as the ponderomotive force resulting from the propagation and variation of amplitude of Alfvén waves \citep[\emph{e.g.}][]{oran_global_2013,gressl_comparative_2013,van_der_holst_alfven_2014}, would potentially solve this problem.
However, we do not want to rely here on physical mechanisms whose inclusion depends on the outcome of the model itself, which is the distribution of fast and slow wind flows.

The goal of this study is to relate the wind speed distributions in Fig. \ref{fig:intro-model-speed} to the geometry of the magnetic features of the low corona shown in Fig. \ref{fig:intro-model-snaps}.
We consider a large sample of magnetic field-lines which probe all open-field zones at different instants of the cycle (hence sampling the full time and latitude intervals).
The field-lines are equally spaced in latitude ($\Delta\theta \approx 1$ degree) at the outer boundary of the numerical domain ($r=15\un{\rsun}$), where they are rooted, and a new set of field-lines is sampled each $0.5$ years.
We then extract a wind velocity profile along each one of them, and perform an \emph{ensemble} analysis in order to correlate the terminal wind speeds with parameters of the correspondent magnetic flux-tubes and to identify trends valid for all latitudes and moments of the cycle.

We do not make direct comparisons with spacecraft data in this manuscript, as our model does not represent real solar dynamo data. 
Our solutions are generically representative of a solar cycle but do not mimic the conditions of one specific cycle.
However, our simulations let us study the evolution of the wind speed and flux-tube expansion ratios much more coherently than what can be achieved by combinations of observational and extrapolation methods, which can be very sensitive to small variations in magnetic connectivity (especially in zones where the magnetic field-lines are very strongly divergent), and systematically produce over and underestimations of the expansion ratios \citep{cohen_quantifying_2015}.
Furthermore, in our model the geometry of the closed-field structures (streamers, pseudo-streamers) result from the interaction of the wind flow with the coronal magnetic field, and so their heights are not limited to a pre-established source-surface height.

\section{Results}
\label{sec:results}

\subsection{Wind speed and expansion}

\begin{figure}[]
  \centering
  \includegraphics[width=\picwd]{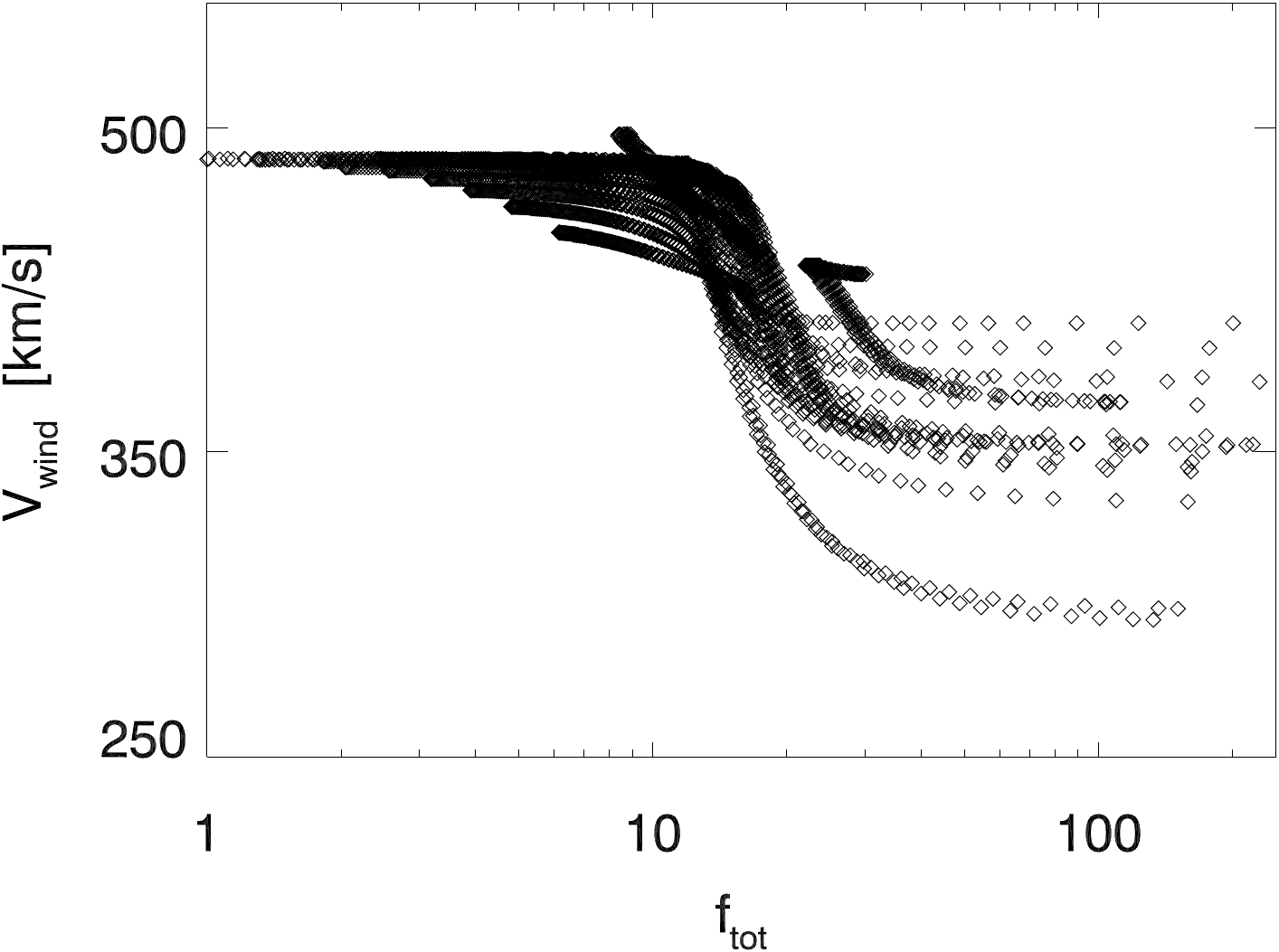}
  \caption{
    Terminal wind speed $V_{wind}$ as a function of the total expansion factor $f_{tot}$ for the whole cycle and for all latitudes.
  }
  \label{fig:f_vs_v}
\end{figure}

\begin{figure}[!h]
  \centering
  \includegraphics[width=\picwd]{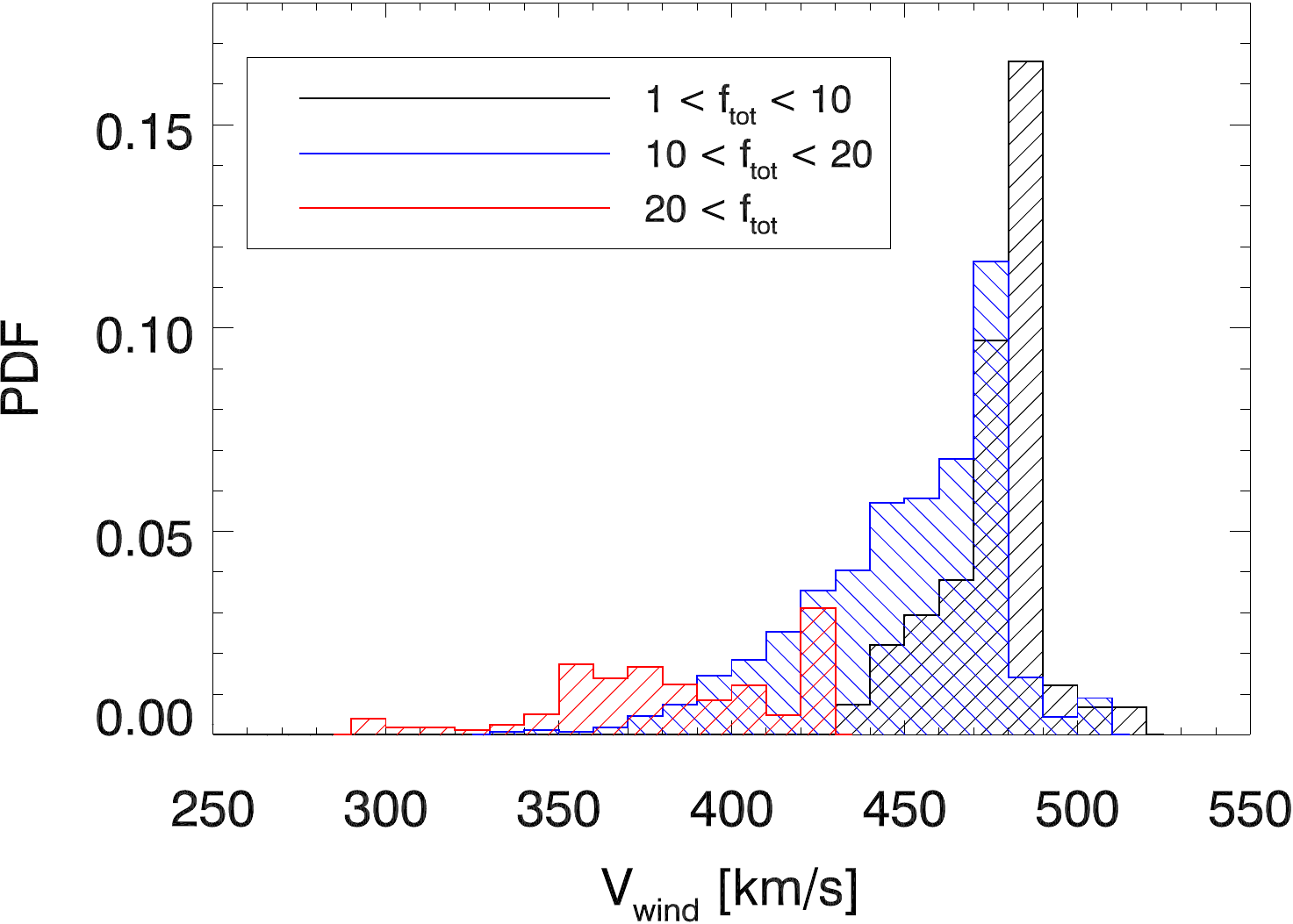}
  \caption{
    Histograms of the terminal wind speed $V_{wind}$ for three different intervals of total expansion factor ($1 \leq f_{tot} < 10$ in black, $10 \leq f_{tot} < 20$ in blue, and $f_{tot} \geq 20$ in red). 
    The data used covers the whole activity cycle and all latitudes.
  }
  \label{fig:hist_f_v}
\end{figure}

Figure \ref{fig:f_vs_v} shows the dependence of the terminal wind speed $V_{wind}$ on the total expansion factor $f_{tot}$ for all points in the dataset (all latitudes and times sampled, altogether).
The total expansion factor is defined as
\begin{equation}
  \label{eq:expans_definition_model}
  f_{tot} = \frac{A_1}{A_0} \left(\frac{r_0}{r_1}\right)^2 = \frac{B_0}{B_1} \left(\frac{r_0}{r_1}\right)^2\ ,
\end{equation}
where $B_0$ and $B_1$ are evaluated respectively at the surface and at the outer boundary of the domain ($15\un{\rsun}$), with the same notation for $A_0$ and $A_1$.
There is, as expected, a general negative correlation between $f_{tot}$ and $V_{wind}$ \citep[\emph{cf.}][]{wang_solar_1990}.
But the terminal wind speed $V_{wind}$ does not seem to be a simple function of $f_{tot}$ alone \citep[see also][]{woolsey_turbulence-driven_2014}.
The spread in the scatter-plot is large, especially for the slow wind part of the diagram, where the data-points are roughly regularly spaced over more than an order of magnitude in $f_{tot}$ but only by a factor $\sim 2$ in $V_{wind}$.
There is furthermore a break in the diagram separating the low wind speed / high expansion part of the diagram from the high speed / low expansion part.

Figure \ref{fig:hist_f_v} shows three histograms of the terminal wind speed for three contiguous intervals of the total expansion factor.
The black bars correspond to the interval $1 \leq f_{tot} < 10$, the blue bars to $10 \leq f_{tot} < 20$, and the red bars to $f_{tot} \geq 20$.
The figure shows that the asymptotic wind speeds are indeed inversely correlated with the expansion factor, but the spreads in the distributions are large and, furthermore, the different distributions overlap each other to a great extent.
That implies that the parameter $f_{tot}$ alone is not enough to determine the terminal wind speed attained on a given flux-tube.

\begin{figure}[!h]
  \centering
  \includegraphics[width=0.73\picwd,clip,trim=0 58 0 0]{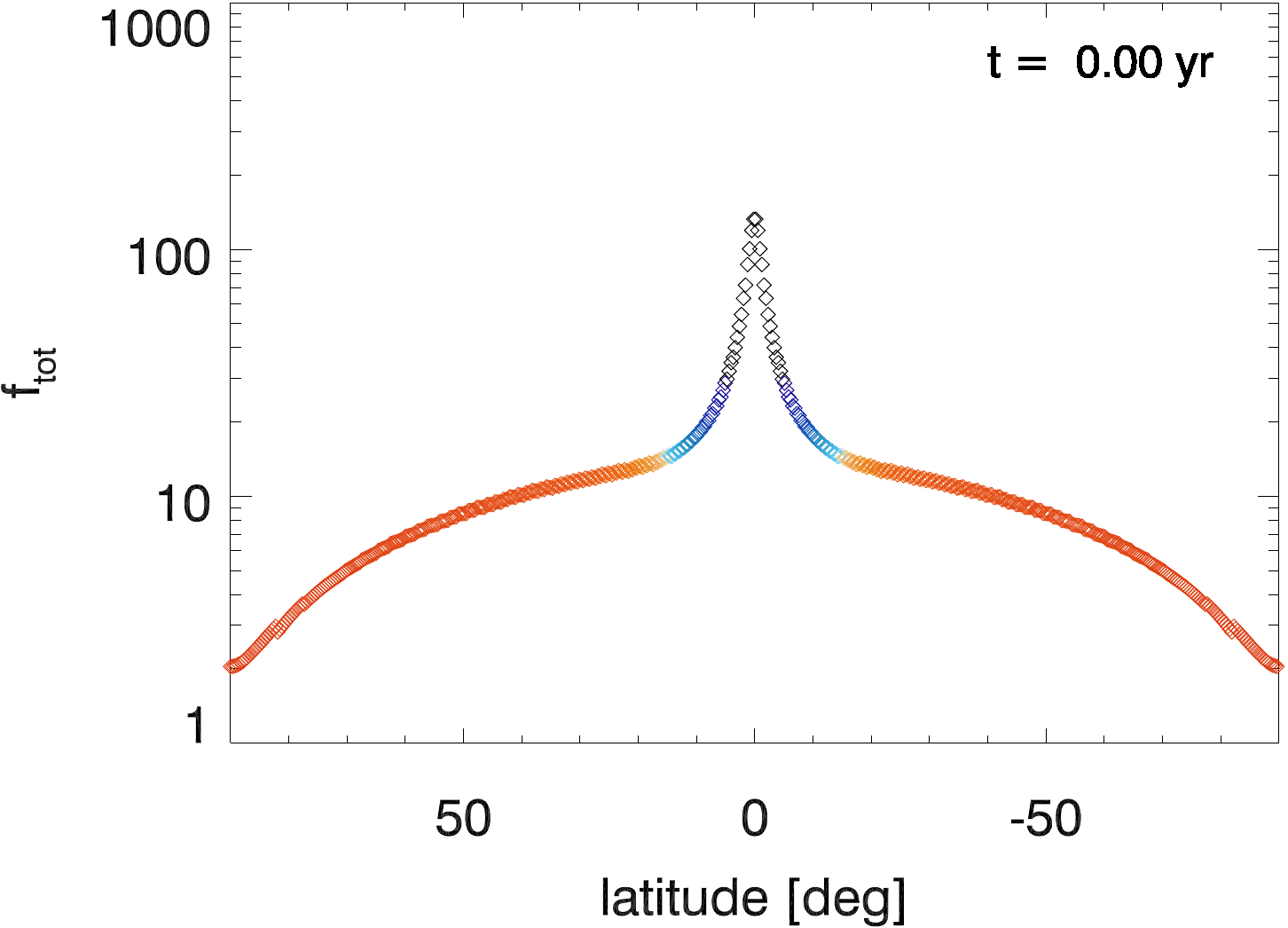}
  \includegraphics[width=0.73\picwd,clip,trim=0 58 0 0]{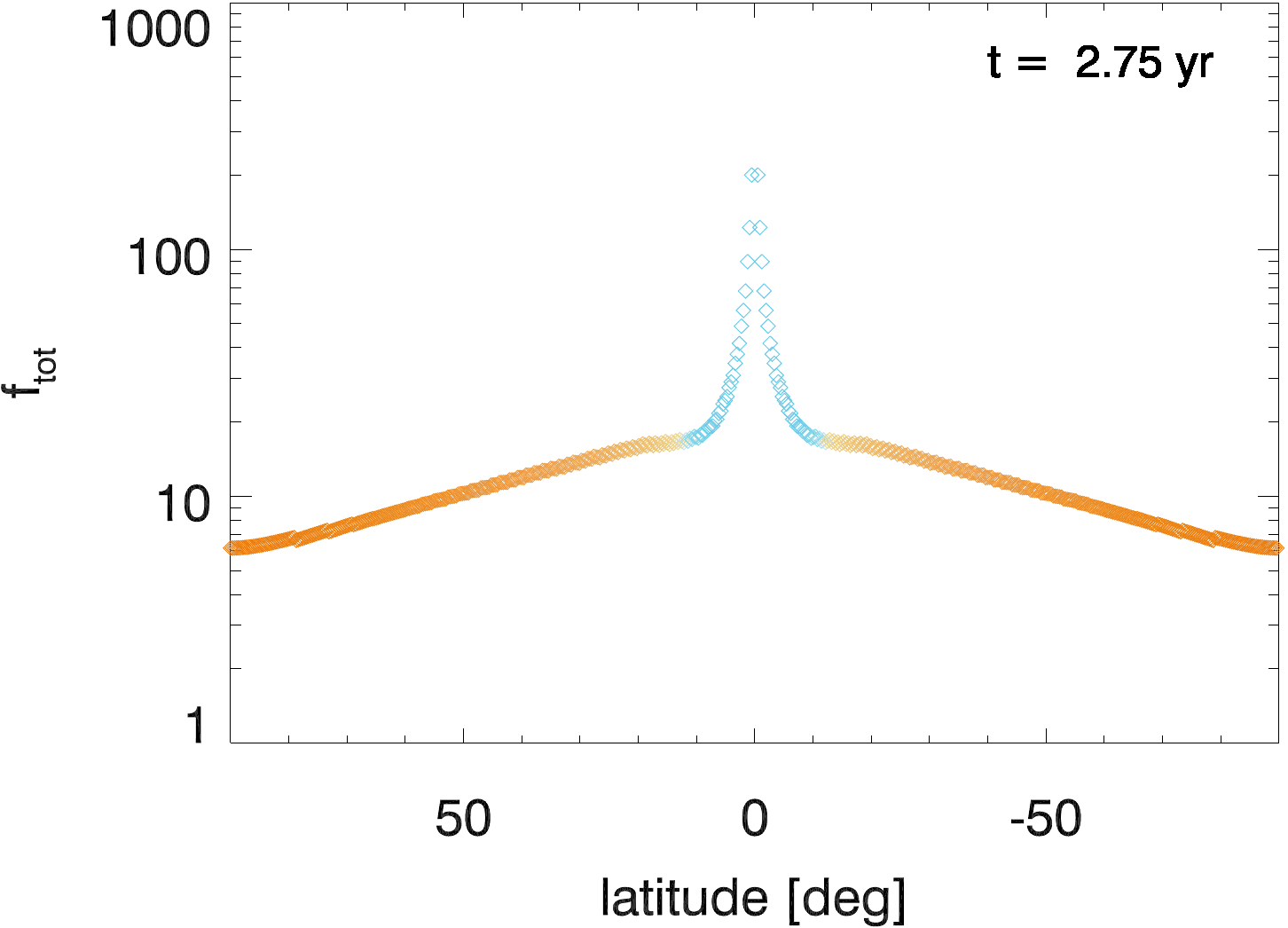}
  \includegraphics[width=0.73\picwd,clip,trim=0 58 0 0]{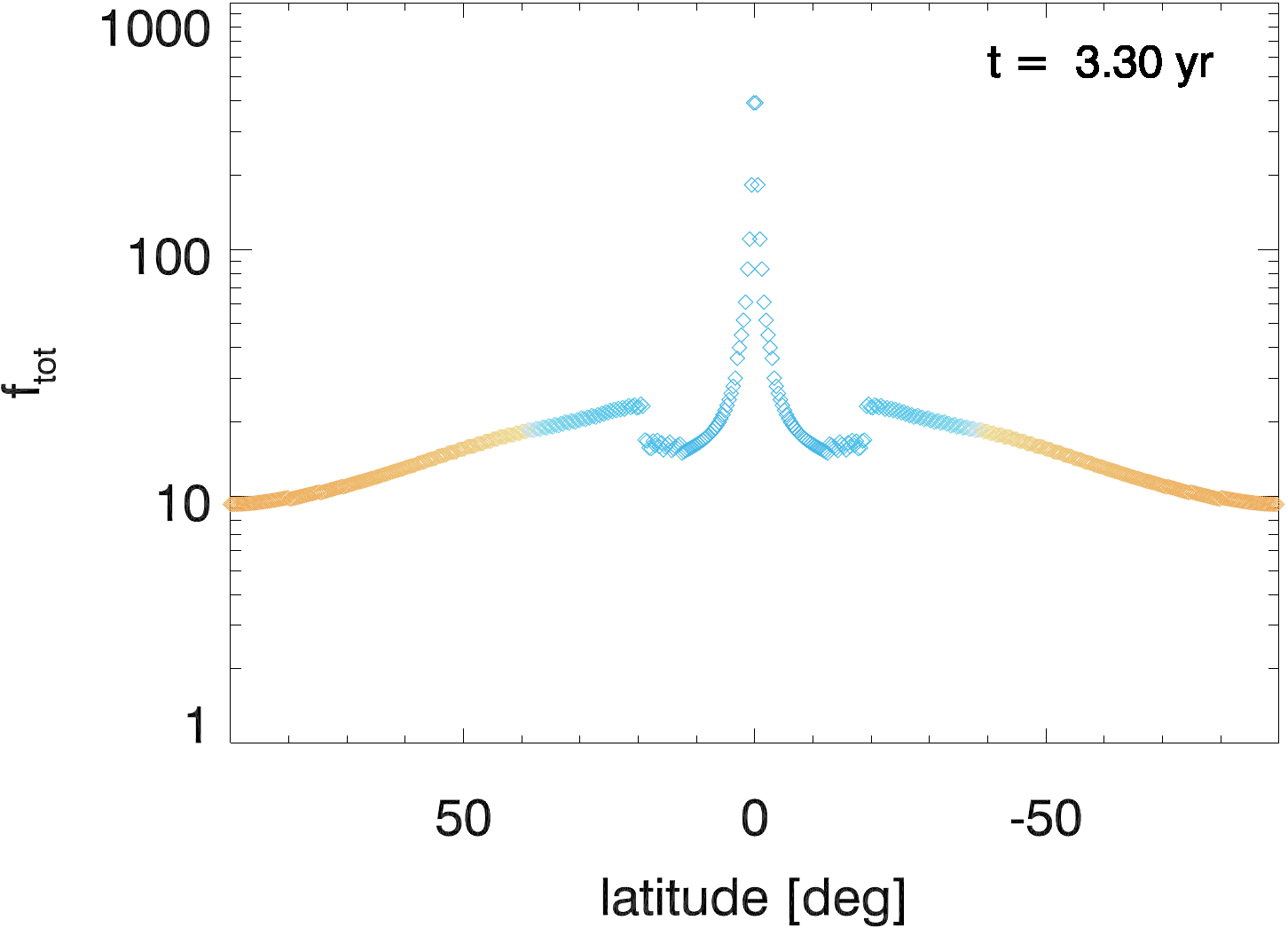}
  \includegraphics[width=0.73\picwd,clip,trim=0 58 0 0]{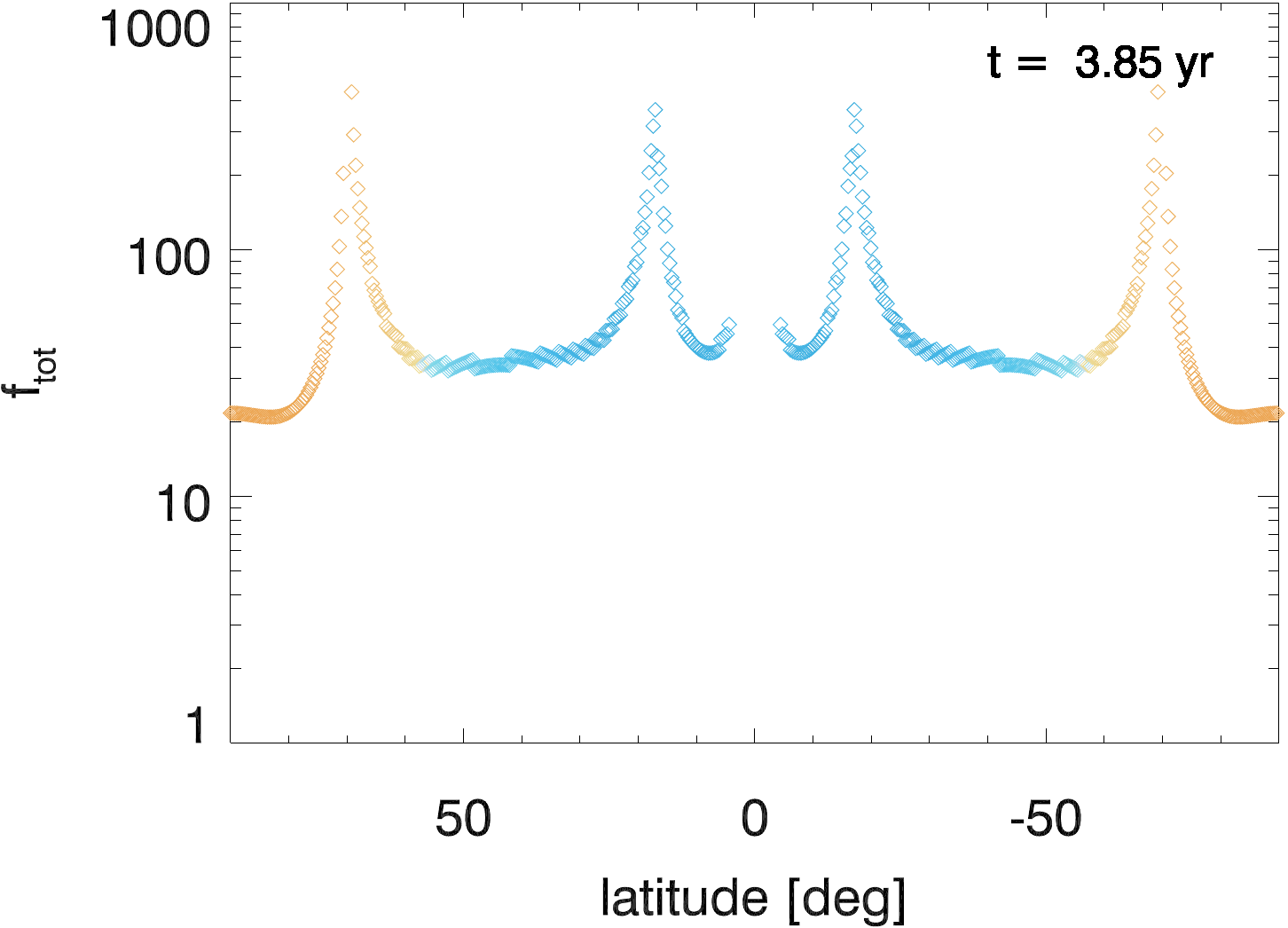}
  \includegraphics[width=0.73\picwd,clip,trim=0  0 0 0]{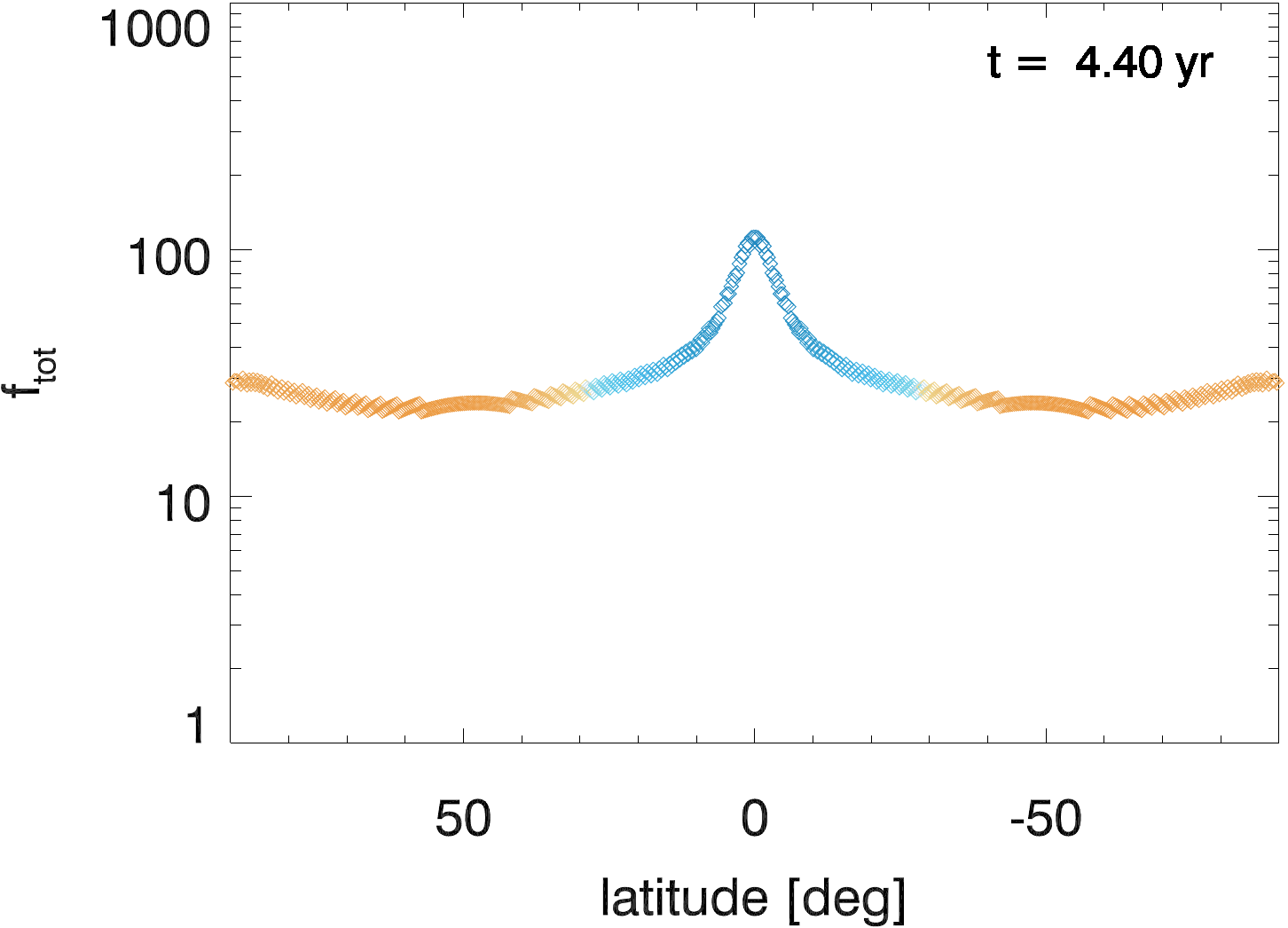}   
  \caption{
    Total expansion factors ($f_{tot}$; see Eq. \ref{eq:expans_definition_model}) as a function of latitude at five instants of the cycle (from the top to the bottom panel: $t=0,\ 2.75\ , 3.30,\ 3.85,\ 4.40 \un{yr}$).
    The field-lines and points are coloured as a function of the terminal wind speed (with the same colour table as in Fig. \ref{fig:intro-model-snaps}).
    The wind speed is anti-correlated with $f_{tot}$ for a large fraction of the simulated activity cycle, but the are exceptions.
    The blue and orange peaks (corresponding to slow and fast wind) in the fourth panel ($t=3.85\un{yr}$) show that flux-tubes with equal values of $f_{tot}$ can bear wind flows with different speeds.
  }
  \label{fig:expans_lat}
\end{figure}

The discrepancy is more evident when comparing smaller subsets of the simulation data.
Figure \ref{fig:expans_lat} represents the total flux-tube expansion factors $f_{tot}$ as a function of field-line latitude (measured at the outer boundary, where the field-lines are rooted) for five subsamples of open magnetic field-lines, corresponding to the five instants represented in Fig. \ref{fig:intro-model-snaps}.
The plot symbols are coloured as a function of terminal wind speed using the same colour-table (dark red for fast wind flows, dark blue for slow wind flows).
During the minimum of activity (first panel), the slow wind flows streams uniquely at the periphery of the coronal holes, while fast wind flows within the wide coronal holes placed at high latitude.
The transition from fast to slow wind streams corresponds to a sharp increase in $f_{tot}$.
In this configuration, the usual semi-empirical hypothesis relating the wind speed with the geometry of the flow applies (at least qualitatively), with the wind speed being strongly dependent on the flux-tube total expansion ratio and/or proximity to the nearest coronal hole boundary (the wind is indeed slower in the flux-tubes with higher expansion rates and/or closer to a S/CH boundary).
However, the situation changes as the activity cycle progress, and becomes very different during the maximum of activity (fourth panel).
The slow wind flows cover a much larger latitudinal extent at the solar maximum, there are many more smaller streamers and current sheets spread in latitude, and the transition between fast and slow winds becomes much smoother.
The total expansion ratios are overall higher than at the minimum ($f_{tot} > 20$ almost everywhere), the angular distances to S/CH boundaries at the surface (parameter $\theta_b$) are much lower (everywhere lower than $1\degree$), but the minimum wind speed is actually higher (by a factor $\sim 1.2$).
Overall, there is one sharp peak in expansion ratio ($f_{tot} \approx 120$) related to a wind speed of about $250\un{km/s}$ at the minimum, and 4 expansion peaks reaching $f_{tot}\approx 400$ corresponding to wind speeds in the range of $350$ -- $400\un{km/s}$.
Our simulations hence suggest that the simple parameters $f_{tot}$ (similar to the parameter $f_{SS}$ found in the literature which makes use of PFSS methods, except in that here expansion can occur for larger height ranges) and $\theta_b$ do not suffice to predict terminal wind speeds accurately.

\begin{figure}[!h]
  \centering
  \includegraphics[width=0.75\picwd,clip,trim=0 60 0 0]{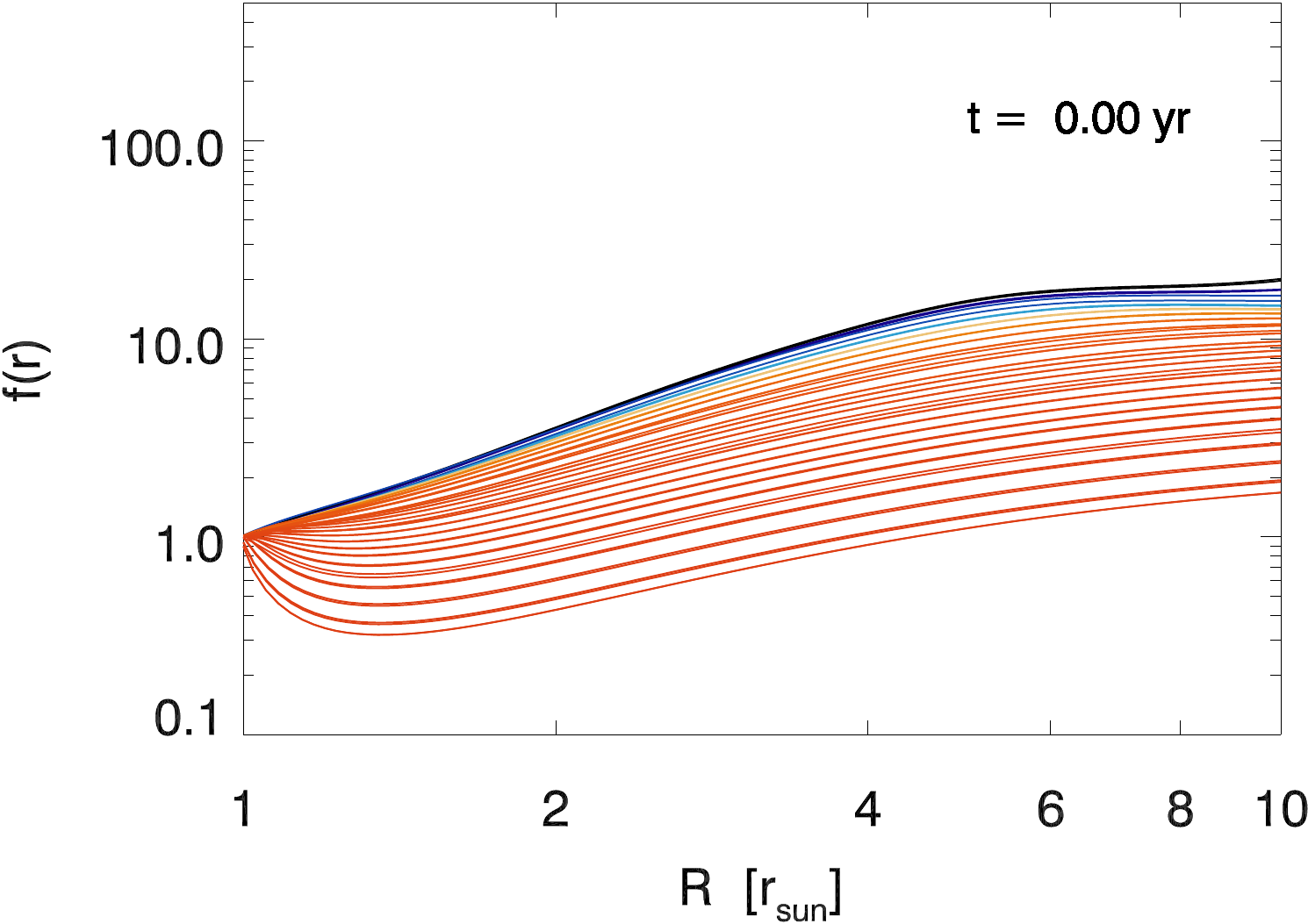}
  \includegraphics[width=0.75\picwd,clip,trim=0 60 0 0]{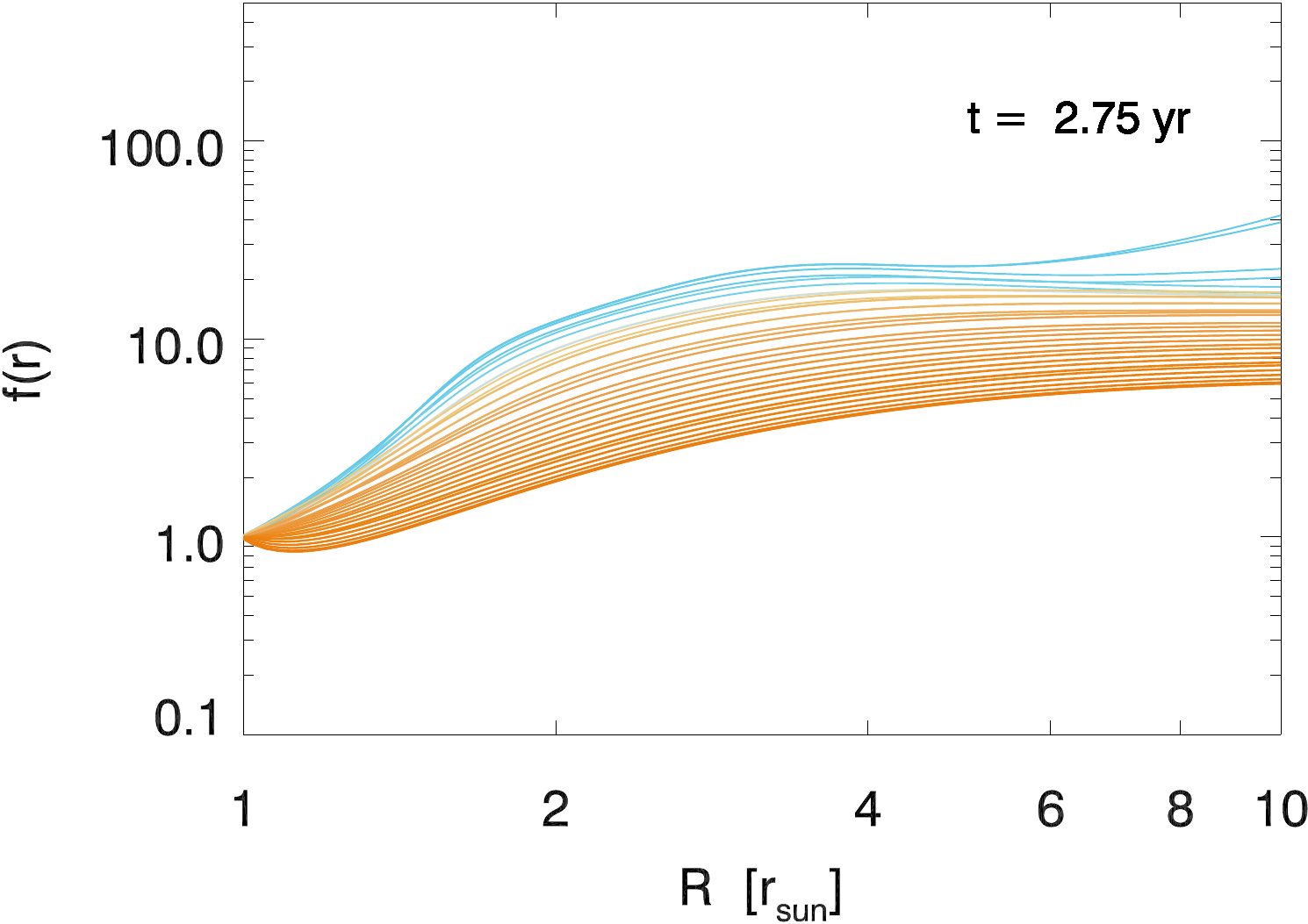}
  \includegraphics[width=0.75\picwd,clip,trim=0 60 0 0]{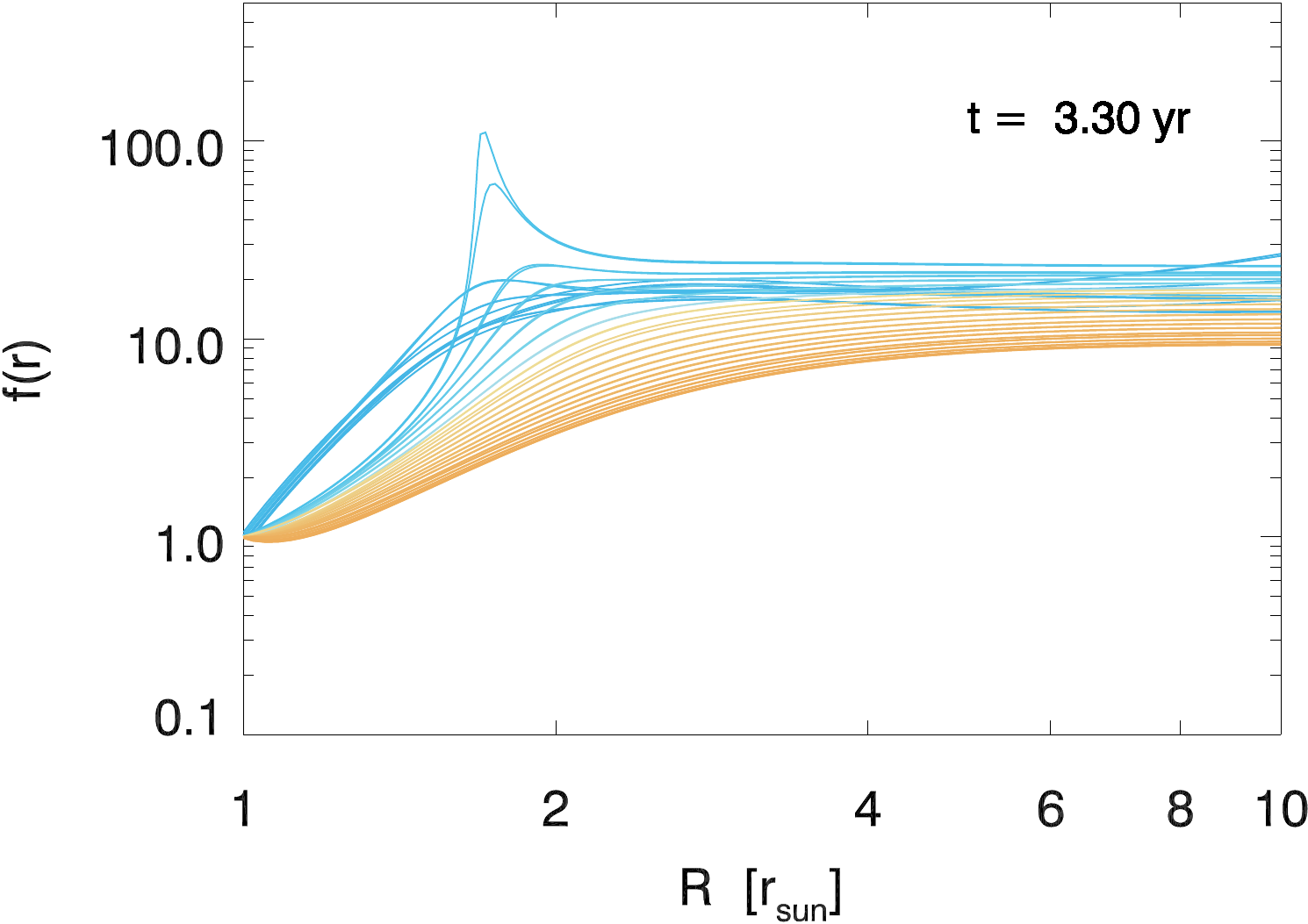}
  \includegraphics[width=0.75\picwd,clip,trim=0 60 0 0]{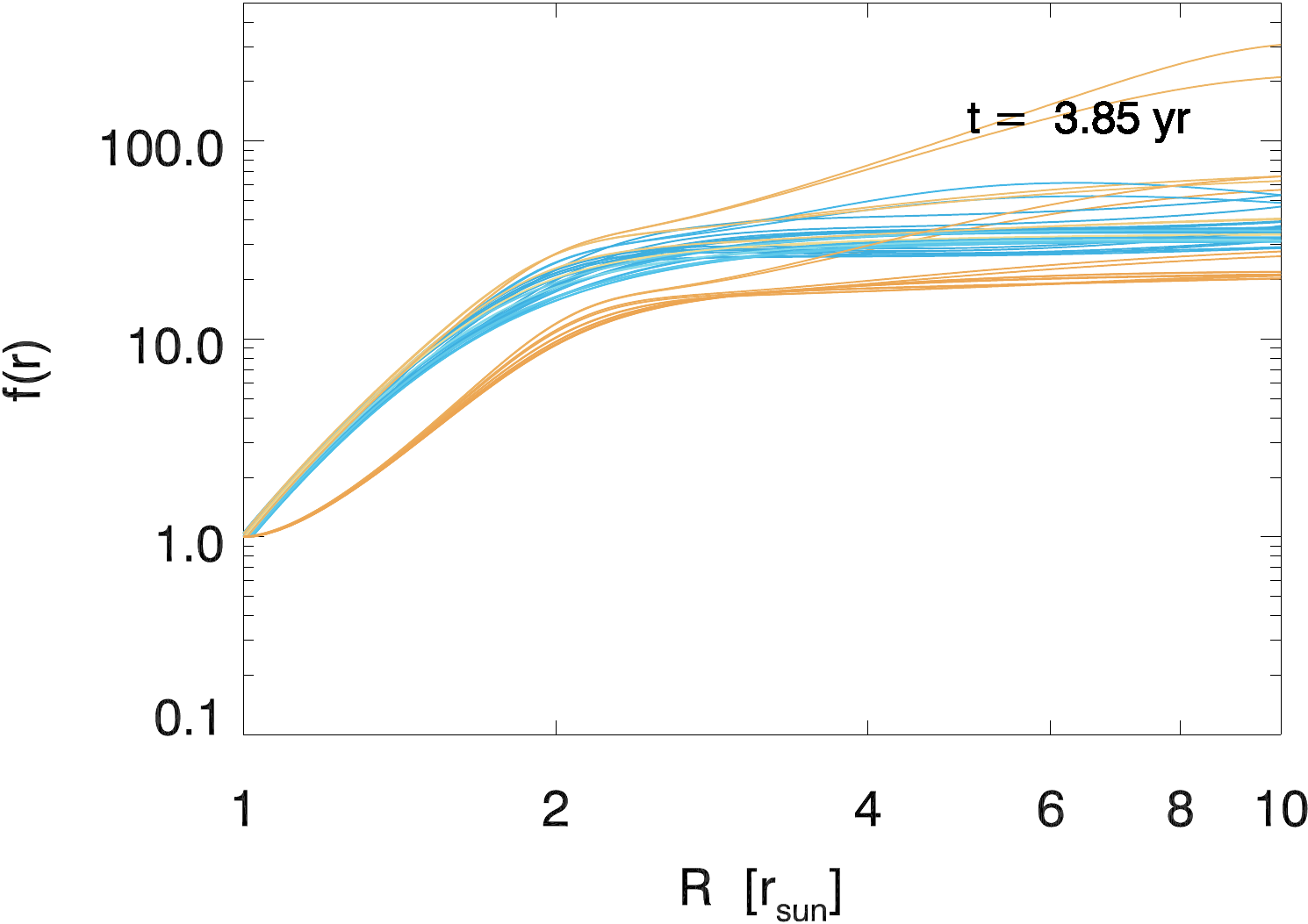}
  \includegraphics[width=0.75\picwd,clip,trim=0  0 0 0]{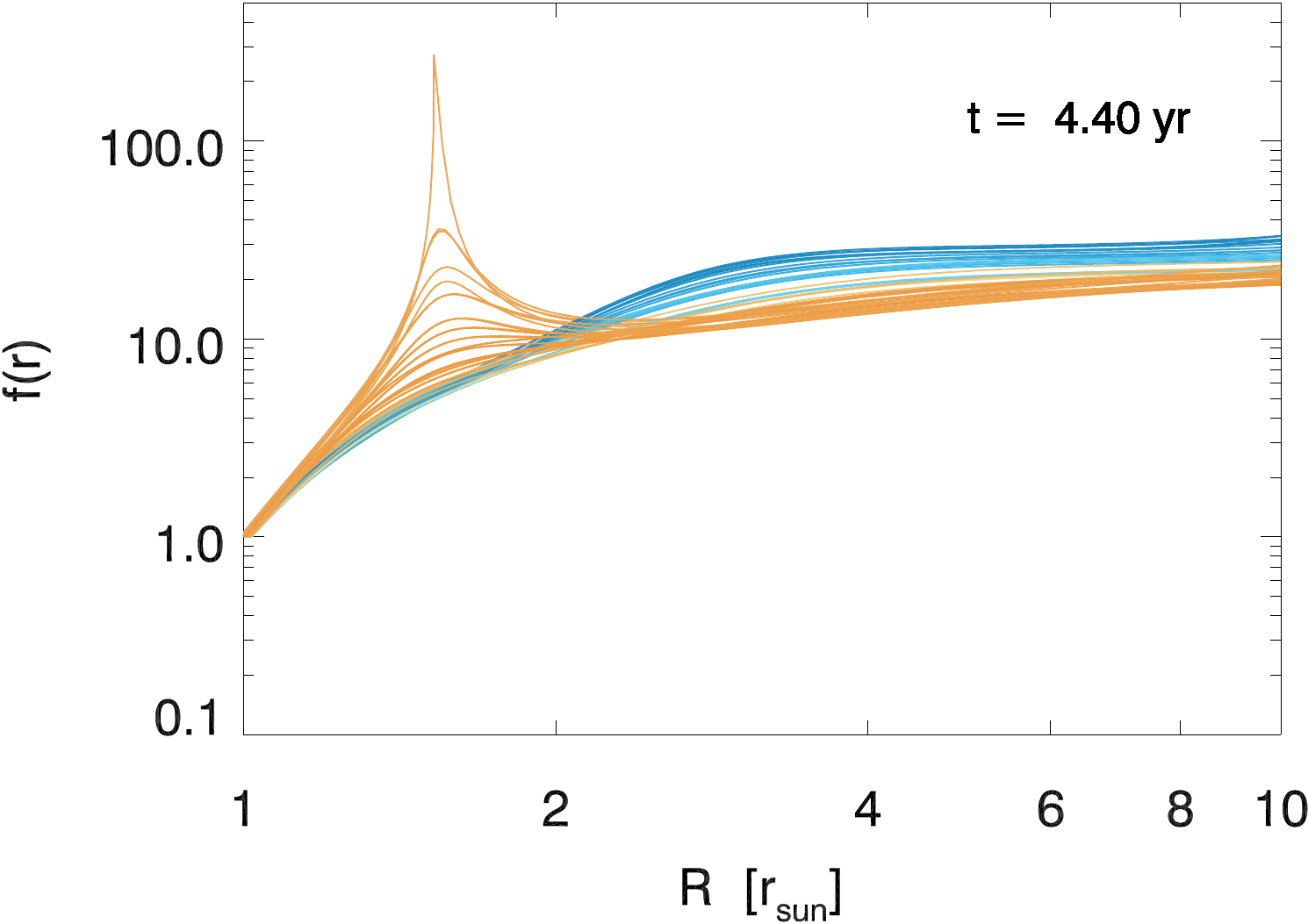}   
  \caption{
    Profiles of the expansion factors as a function of the distance to the surface $f\left(r\right)$ corresponding to flux-tubes at different latitudes at five different instants (from the top to the bottom panel: $t=0,\ 2.75\ , 3.30,\ 3.85,\ 4.40 \un{yr}$).
    The curves are coloured according to the asymptotic wind speed, as in the previous figures.
    The figures show that the terminal wind speed is anti-correlated with the total expansion factor (the value of $f\left(r\right)$ at $r \gg 1\un{\rsun}$) -- especially close to the solar minimum --, but not so much to the details of $f\left(r\right)$ at low altitudes.
  }
  \label{fig:expans-distance}
\end{figure}

\begin{figure}[]
  \centering
  \includegraphics[width=\picwd]{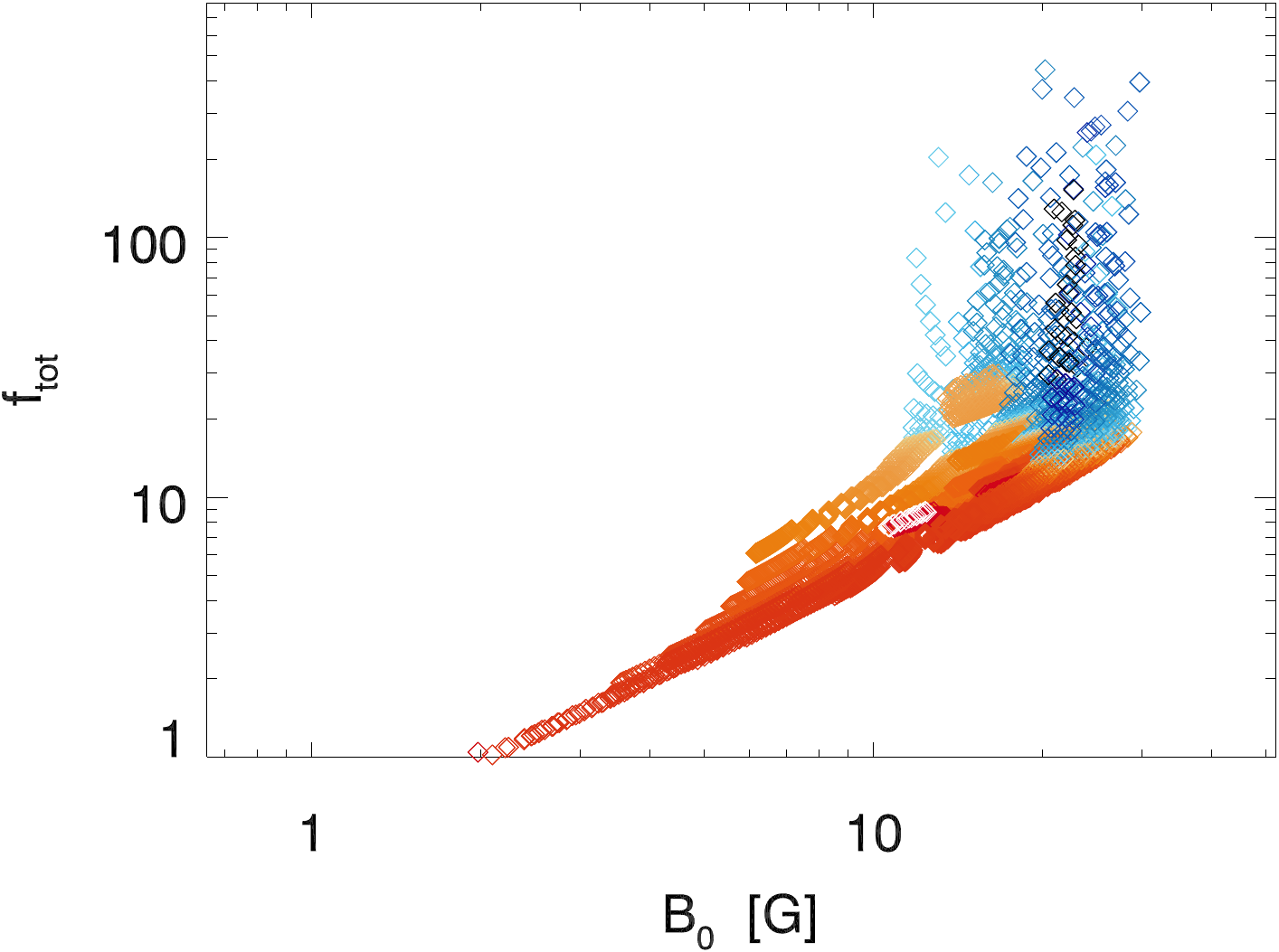} \\
  \includegraphics[width=\picwd]{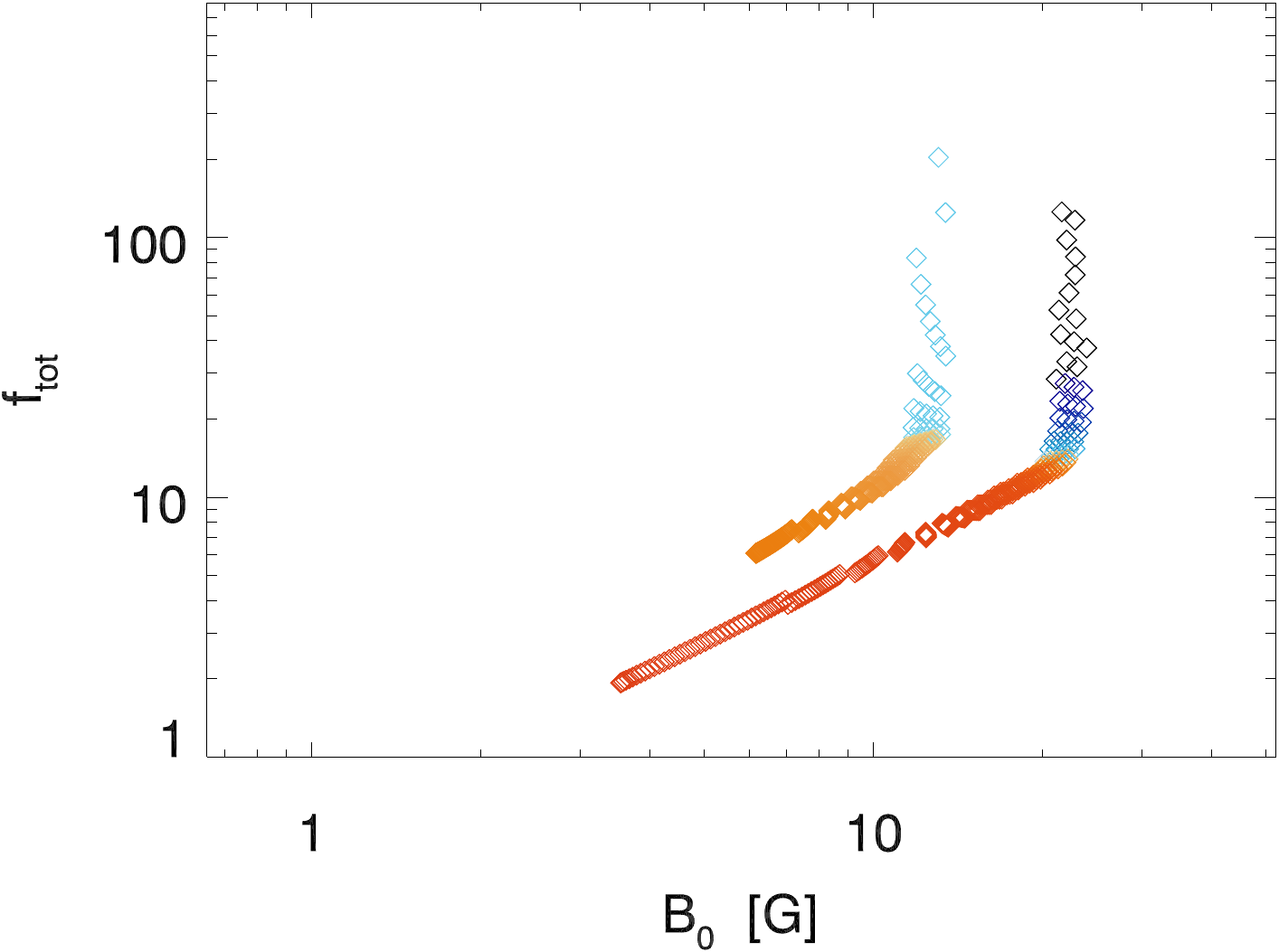}   
  \caption{
    Scatter-plots of the total flux-tube expansion rate as a function of the magnetic field strength at the surface.
    The top panel represent these quantities calculated from our simulations for an $11\un{yr}$ period.
    The points are colour-coded according to the speed attained by the wind flow in each flux-tube (orange represents fast wind, blue represents the slow wind).
    The bottom panel represents only the simulation data-points relative to the minimum and to the maximum (first and third panels in Fig. \ref{fig:expans_lat}).
  }
  \label{fig:scatter-f-b0}
\end{figure}
\begin{figure}[]
  \centering
  \includegraphics[width=\picwd]{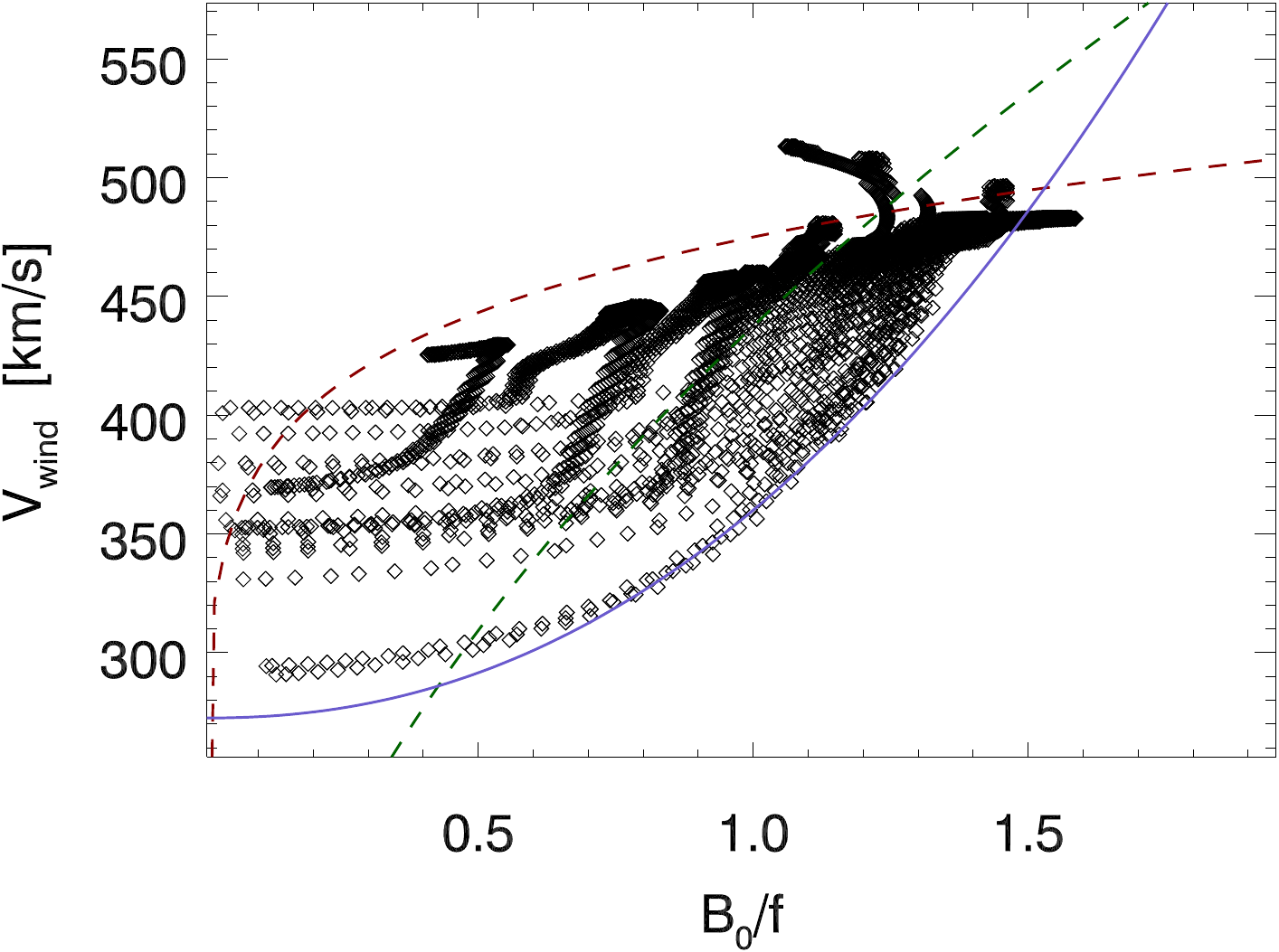}
  \caption{
    Terminal wind speed $V_{wind}$ as a function of the ratio $B_0/f_{tot}$ for the full data-set.
    The green dashed line corresponds to the curve $\left(B_0/f_{tot}\right)^{0.5}$ proposed by \citet{suzuki_forecasting_2006} normalised to an arbitrary coefficient for easier representation.
    The brown dashed and blue continuous lines represent the curves $\left(B_0/f_{tot}\right)^{0.1}$ and $\left(B_0/f_{tot}\right)^{2.2}$, which are empirical fits to the upper and lower envelopes of the data points (correspondingly, fast and slow wind limits; the curves are also normalised to arbitrary coefficients). 
  }
  \label{fig:scatter-v-b0f}
\end{figure}
\begin{figure}[!h]
  \centering
  \includegraphics[width=0.75\picwd,clip,trim=0 60 0 0]{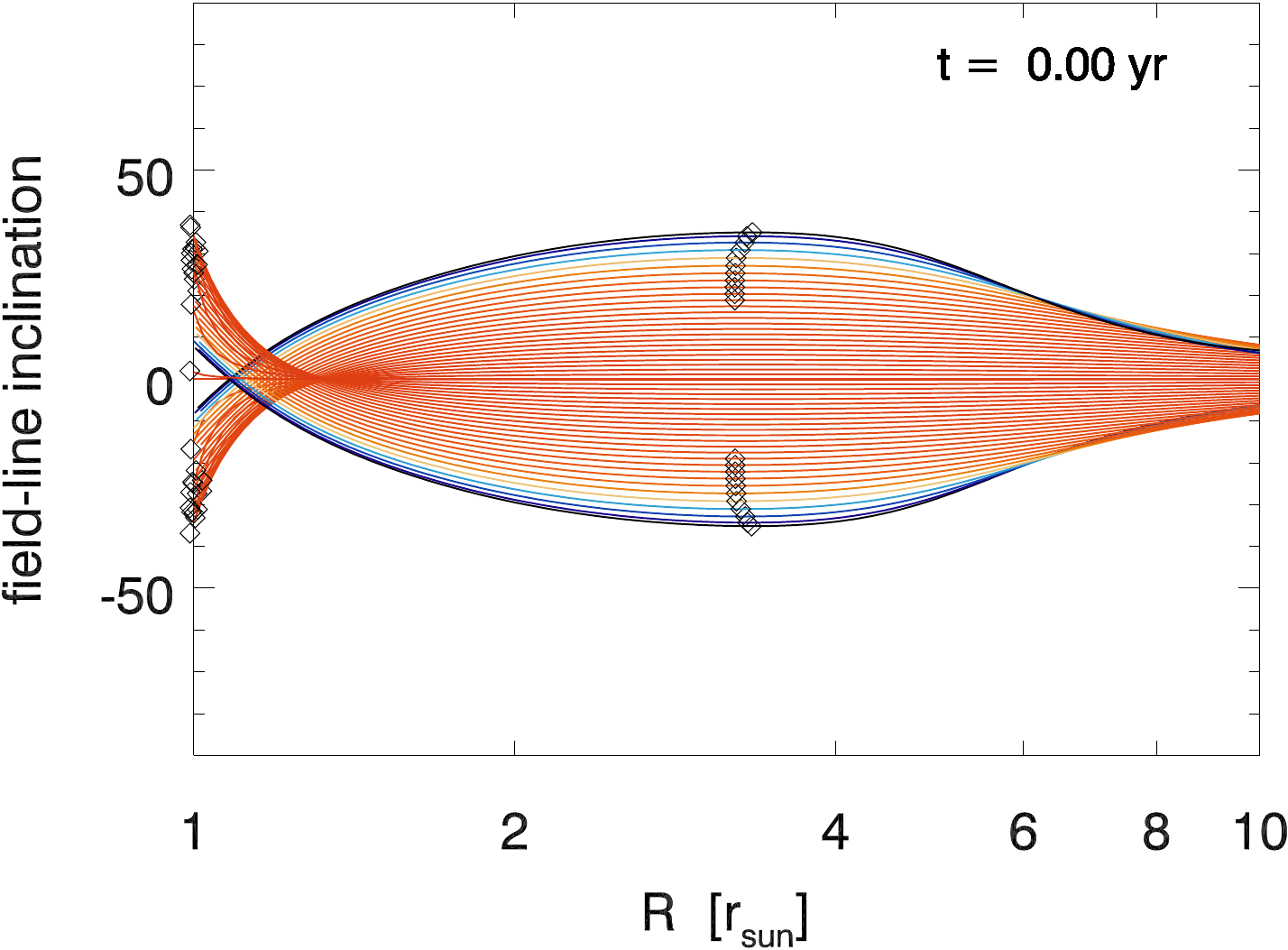}
  \includegraphics[width=0.75\picwd,clip,trim=0 60 0 0]{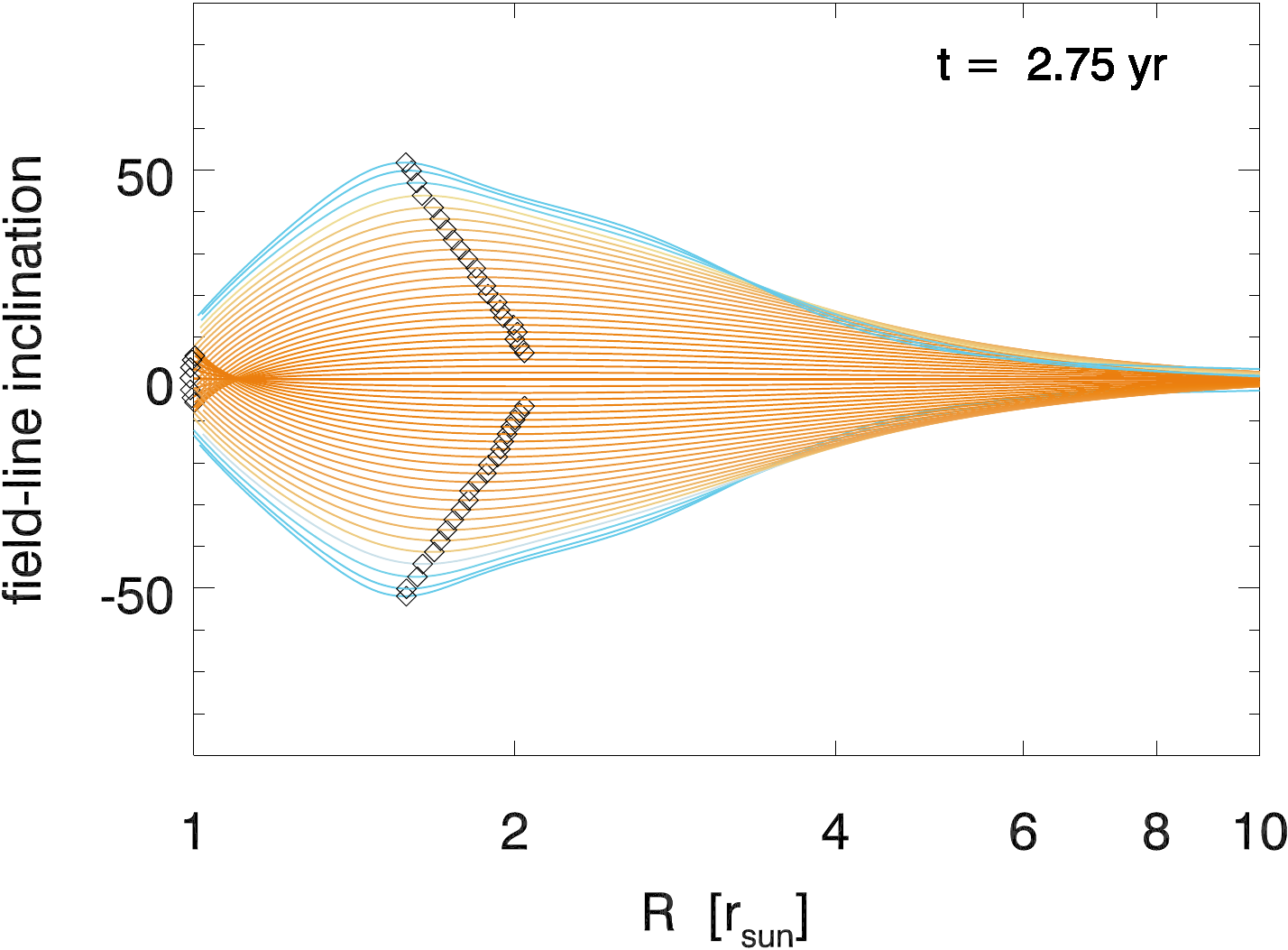}
  \includegraphics[width=0.75\picwd,clip,trim=0 60 0 0]{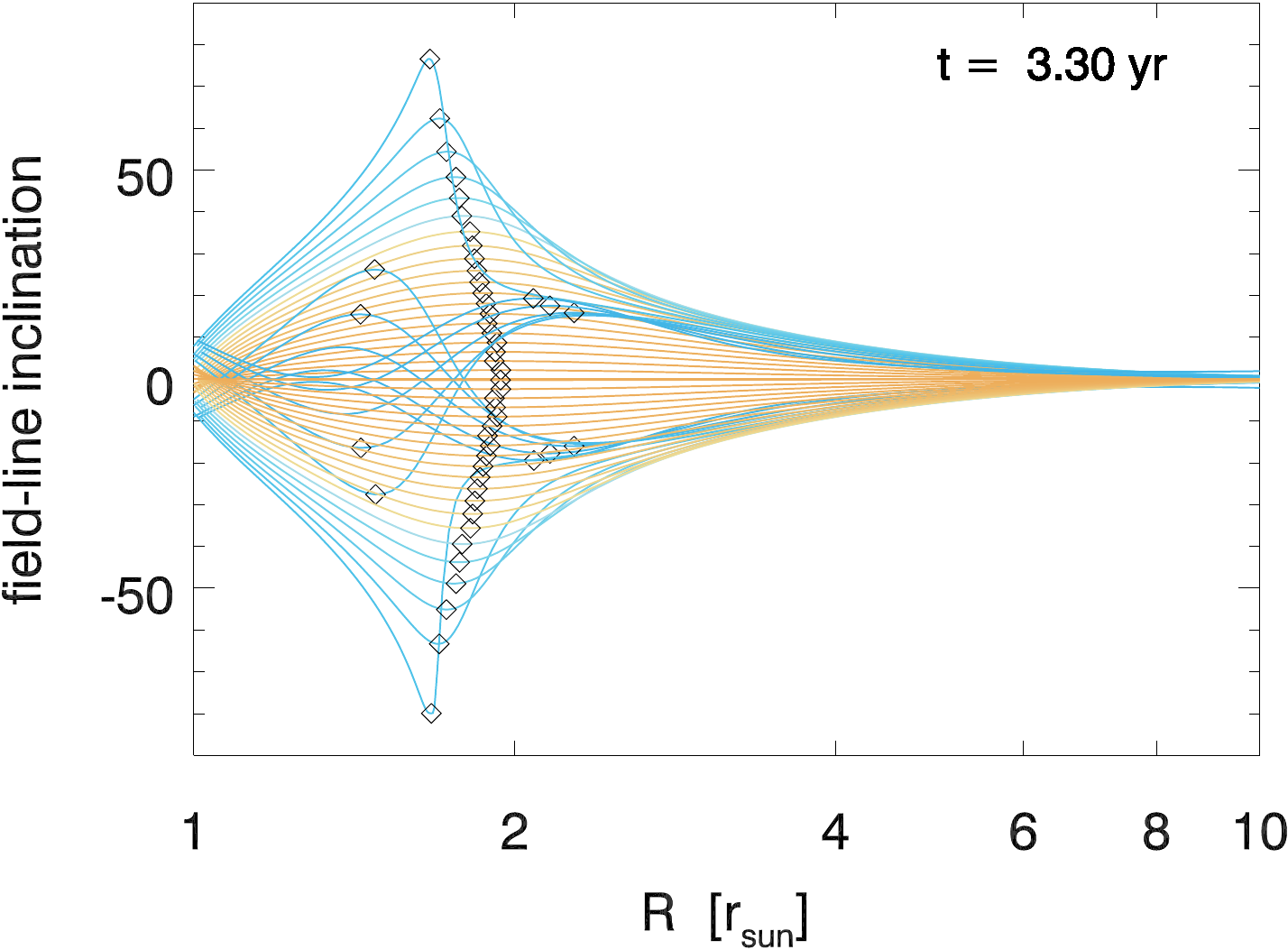}
  \includegraphics[width=0.75\picwd,clip,trim=0 60 0 0]{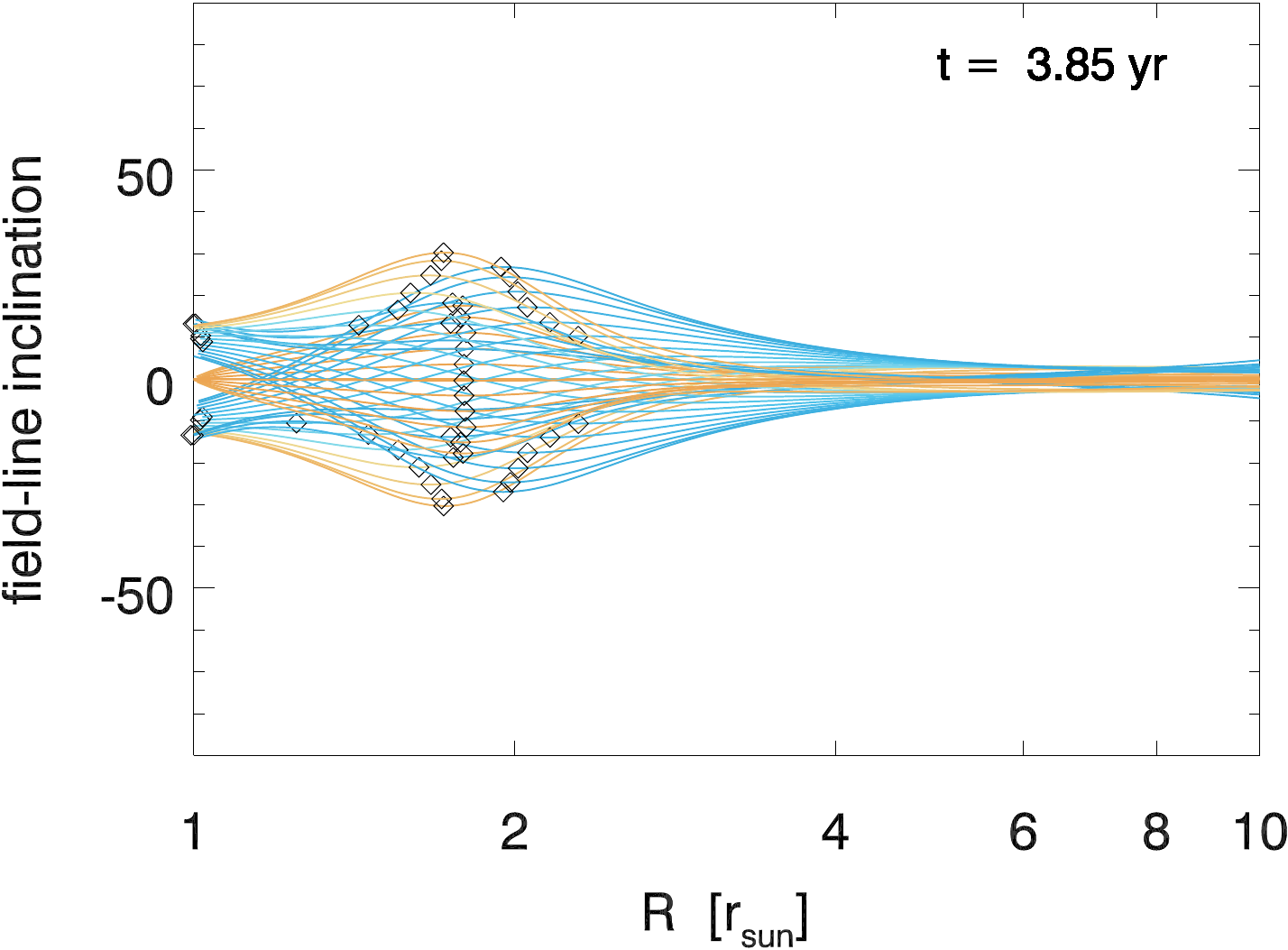}
  \includegraphics[width=0.75\picwd,clip,trim=0  0 0 0]{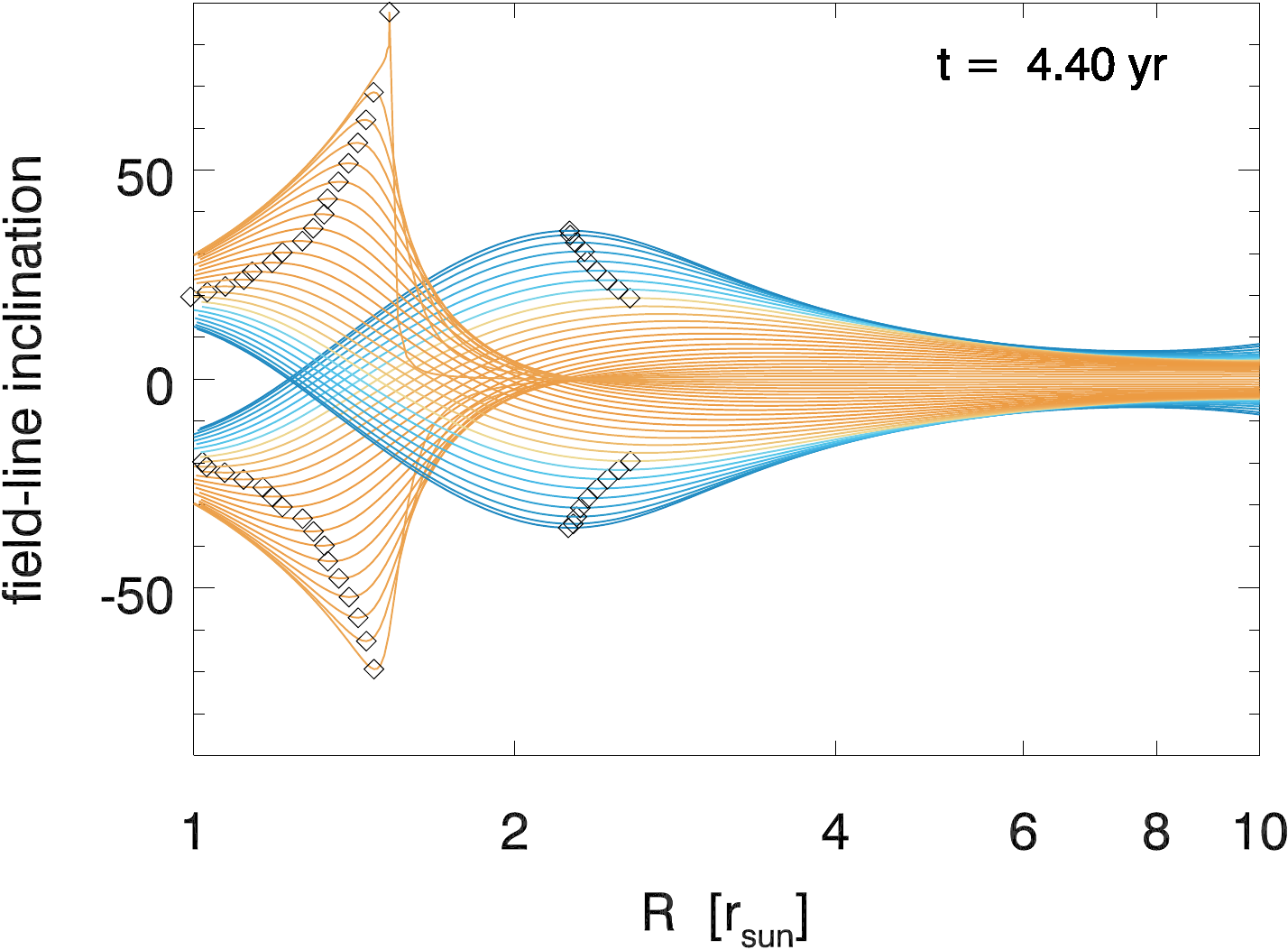}   
  \caption{
    Profiles of the field-line inclination angle (angle to the vertical direction, in degrees, from $-90$ to $90$) as a function of the distance to the surface for flux-tubes at different latitudes and at five different instants of the cycles (from the top to the bottom panel: $t=0,\ 2.75\ , 3.30,\ 3.85,\ 4.40 \un{yr}$).
    The diamonds mark the position of the maximum inclination for each flux-tube.
    The times and colour scheme are the same as in Figs. \ref{fig:expans_lat} and \ref{fig:expans-distance}.
  }
  \label{fig:profiles-inclin}
\end{figure}

\begin{figure}[!h]
  \centering
  \includegraphics[width=\picwd]{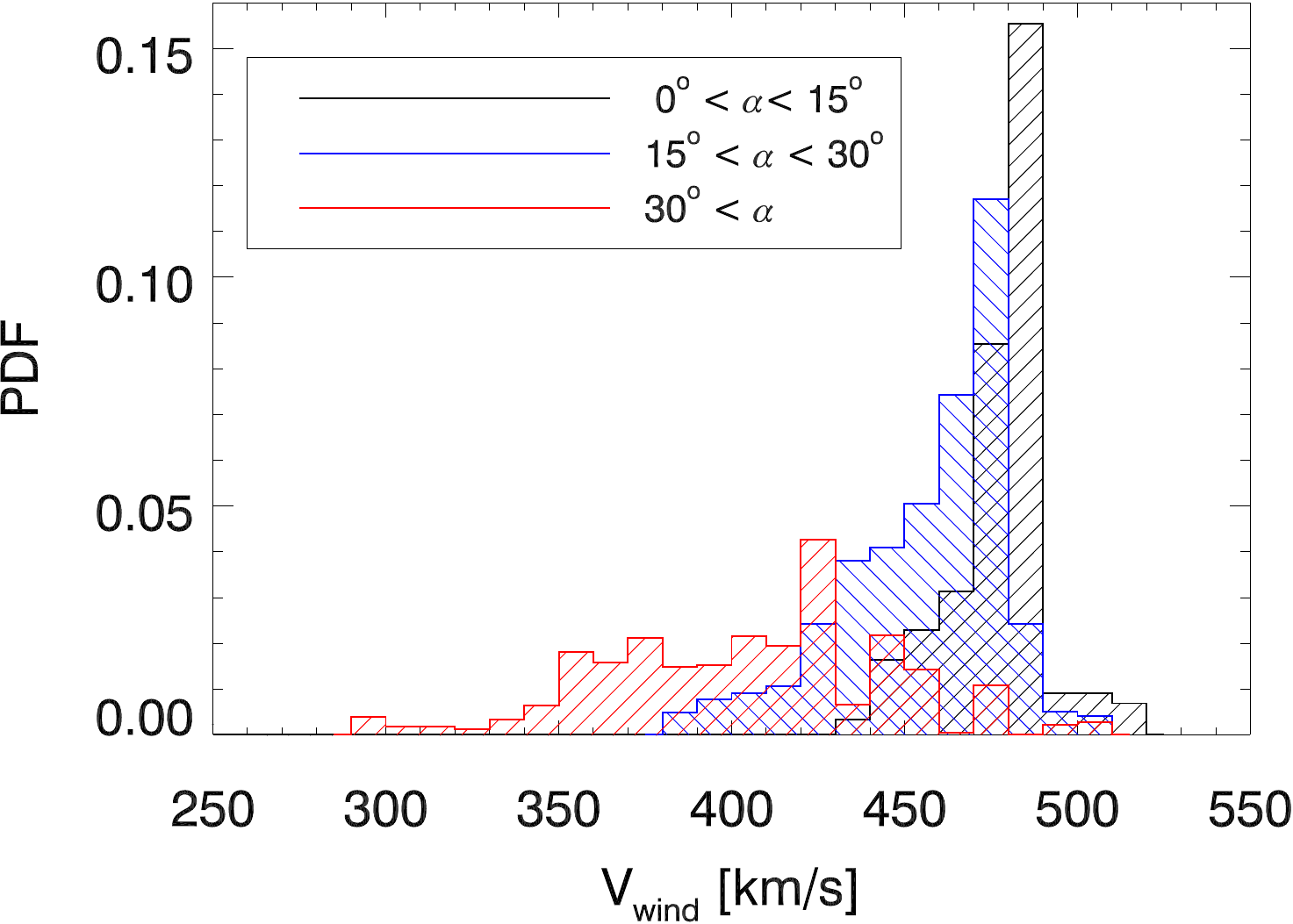}
  \caption{
    Histograms of the terminal wind speed $V_{wind}$ for three different intervals of the maximum field-line inclination ($0^\circ \leq \alpha < 15^\circ$ in black, $15^\circ \leq \alpha < 30^\circ$ in blue, and $\alpha \geq 30^\circ$ in red). 
    The data used covers the whole activity cycle and all latitudes.
  }
  \label{fig:hist_i_v}
\end{figure}

\begin{figure}[!h]
  \centering
  \includegraphics[width=\picwd]{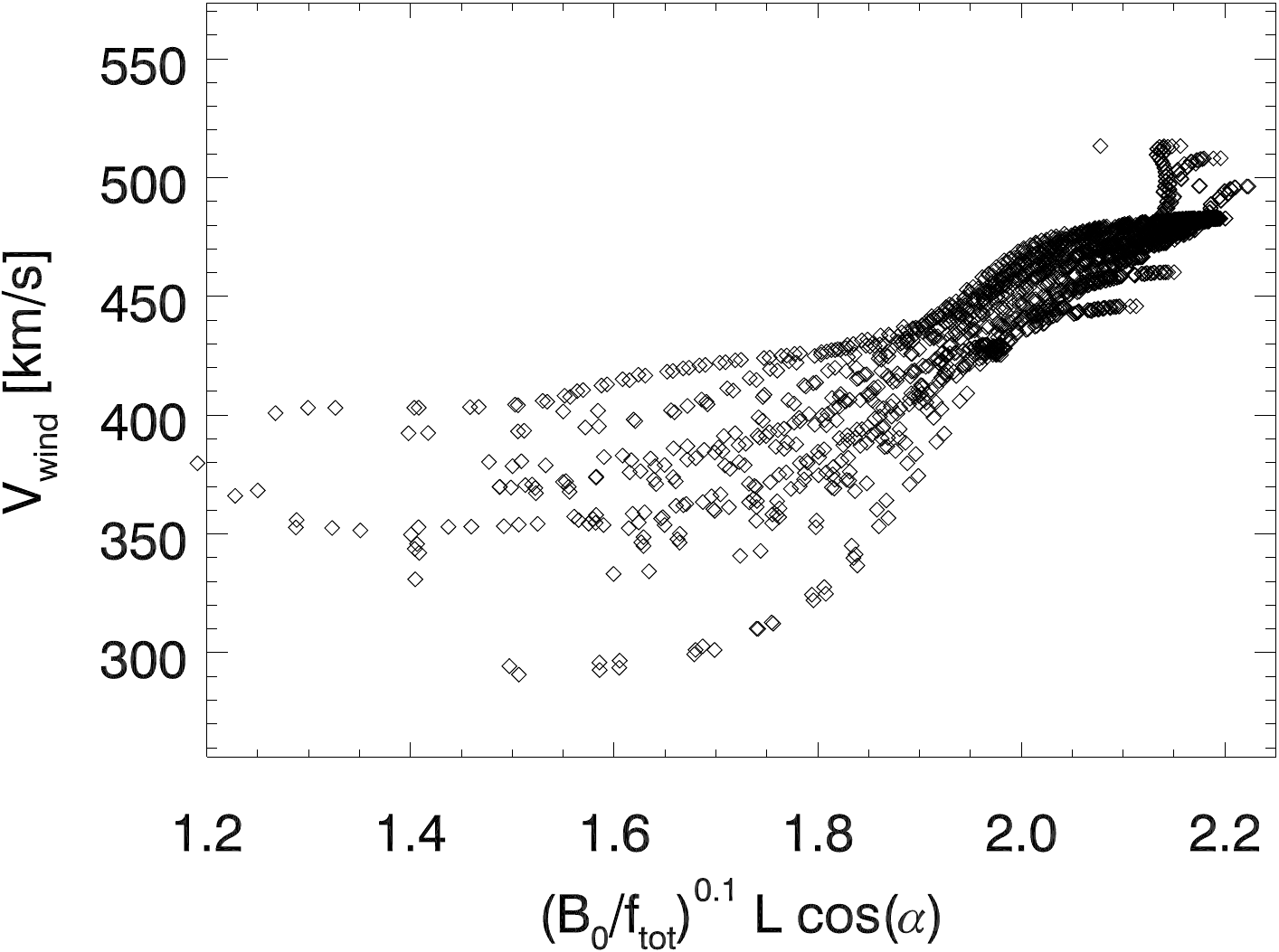}
  \caption{Terminal wind speed $V_{wind}$ as a function of $\left(B_0/f_{tot}\right)^{0.1} \times L \cos\left(\mathrm{\alpha}\right)$, where $L$ is the field-line length, and $\alpha$ is the maximum inclination of each field-line in respect to the vertical direction.
  }
  \label{fig:speed-inclin}
\end{figure}

Let us now examine the expansion profiles of our set of flux-tubes, that is, the way the cross-section of each flux-tube varies with height.
Figure \ref{fig:expans-distance} represents several expansion profiles
\begin{equation}
  \label{eq:expans_profile}
  f\left(l\right) = \frac{A\left(l\right)}{A_0} \left(\frac{r_0}{l}\right)^2 = \frac{B_0}{B\left(l\right)} \left(\frac{r_0}{l}\right)^2\ ,
\end{equation}
where $l$ is the distance to the surface measured along the field-line, such that $f_{tot} = f\left(l \gg\rsun\right)$.
The curves are colour-coded as those in the preceding figures, with orange representing flows with high terminal speed and blue representing slow wind flows.
The figure shows that most of the flux-tubes display smooth and steadily increasing expansion profiles, especially during the minimum of the activity cycle.
Conversely, flux-tubes that cross the vicinity of small streamers and pseudo-streamers (more common during the rising phase and the maximum of the activity cycle) often display a sharp increase in $f\left(l\right)$ followed by an abrupt inflexion at low coronal altitudes, which then evolve into a more conventional and smoother over-expansion \citep[\emph{cf.}][]{wang_nature_2012}.
The terminal speeds of the wind flows in these particular flux-tubes is in most cases related to the total expansion ($f\left(l\right)$ faraway from the surface), rather to the maximum values of  $f\left(l\right)$ occurring below the height of the streamers (well visible in the third and fifth panels of Fig. \ref{fig:expans-distance}).
In fact, flux-tubes showing such strong inflexions can produce both slow and fast wind flows, as shown in the third and fifth panel of Fig. \ref{fig:expans-distance}.
It would be physically sound to search for an additional parameter describing the radial expansion of the flux-tubes (as the height or height interval over which strong expansion occurs rather than just the total expansion), but our results do not reveal the effect any such parameter clearly and unambiguously (even though these features seem at first to be more often related to fast wind flows when they occur lower down in the corona, and to slower flows when they occur higher up in our simulations).

\subsection{Wind speed, expansion and magnetic field amplitude}



Other predictive theories for the solar wind speed combine the magnetic field amplitude at the foot-point of the flux-tubes in addition to their expansion ratios \citep[\emph{e.g}][]{kovalenko_origin_1978,kovalenko_energy_1981,wang_why_1991,suzuki_forecasting_2006}.
This is justified, in part, by considering that the energy input to the solar wind takes the form of a Poynting flux resulting from horizontal surface motions at the surface of the Sun acting on the magnetic field lines which cross it and extend into the corona.
The corresponding energy flux density (the Poynting flux) can in that case be expressed as $B_0 v_\perp \sqrt{\rho}$ at the surface, where $B_0$ is the amplitude of the magnetic field there, $v_\perp$ is the amplitude of the transverse motions and $\rho$ is the mass density.
The factor $B_0$ is the most significant one in the expression, as long as it is assumed that $v_\perp \sqrt{\rho}$ has the same statistical properties across the whole solar surface.
\citet{suzuki_forecasting_2006}, in particular, suggests that the wind speed should scale as $V_{wind} \propto \sqrt{B_0 / f_{tot}}$ (keeping the coronal temperature and the wave power input constant).
The amplitude of the magnetic field and the way it decays with height can also influence the rate and position at which this energy input is dissipated in the corona, on which its effect on the terminal wind speed depends \citep[see \emph{e.g.}][ and many others]{pinto_time-dependent_2009,cranmer_coronal_2002}.
Alternatively, one can interpret this relationship between wind speed and $B_0 / f_{tot}$ as resulting from a conjugation of geometrical properties of the coronal magnetic field.
It is easy to see from Eq. (\ref{eq:expans_definition_model}) that this ratio is an expression of the magnetic field amplitude away from the Sun (or of the open magnetic flux), whose average value is known to vary cyclically, being smaller near the solar minima and larger near the solar maxima.
\cite{fujiki_relationship_2015} have, in fact, shown that the degree of correlation between $B_0/f$ and wind speed itself varies systematically over the solar cycle using a combination of PFSS extrapolations and IPS radio data.
They attributed this to the fact that the magnetic field amplitude at the source-surface and the average wind speed vary both during the solar cycle, albeit with different relative variations (and possibly different causes).

Figure \ref{fig:scatter-f-b0} shows the total flux-tube expansion rate $f_{tot}$ plotted against the foot-point magnetic field amplitudes $B_0$ in our simulations.
Each point in the scatter-plot corresponds to one of the flux-tubes sampled.
The points are coloured using the same colour scheme as in Figs. \ref{fig:expans_lat} and \ref{fig:expans-distance}: red corresponds to high terminal wind speeds and blue to low terminal wind speeds.
The top panel shows the whole set of flux-tubes, while the bottom panel shows only data-points for instants relative to the minimum and to the maximum (at the same two instants as in Fig. \ref{fig:expans_lat}).
%
The main features of the full $f_{tot} - B_0$ diagram (top panel) are a general positive trend/slope (although with considerable scatter), a well defined cut-off at the low-end of the scatter plot, a sharp transition from fast to slow wind regimes (more clearly visible on the second panel), and a dependence of the wind speed on both quantities.
The $f_{tot} - B_0$ curves shown on the bottom panel show that, at a given moment of the cycle, these two quantities relate to one another one and follow a broken power-law of the kind $f_{tot} \propto B_0^\alpha$, with different values of the index $\alpha$ for the slow and for the fast wind.
The fast wind regime corresponds to a moderate $\alpha$, while the slow wind regime corresponds to a much steeper (and almost undefined) index $\alpha$.
These $\alpha$ indexes are roughly invariant throughout all the cycle.
The strongest deviation occurs when the background magnetic field is at its most multipolar state, at about the maximum of activity.
The spread in the $f_{tot} - B_0$ diagram increases when all instants of the cycle are represented, but the properties described above are maintained throughout.
The ``low-end cutoff'' of the $f_{tot} - B_0$ diagram corresponds to the configuration of the coronal magnetic field at the solar minimum, when it reaches its simpler topological configuration (\emph{i.e} quasi-dipolar).
This is the state at which the smallest expansion ratios are attained ($f_{tot}\approx 1$, close to the poles).
The minimal expansion ratios progressively increase as the cycle proceeds towards the maximum of activity.
Equivalently, the latitudinal extent of the coronal holes at the surface decreases from minimum to maximum.
The largest deviation with respect to the ``low-end cutoff'' line corresponds to the solar maximum.
The overall red -- blue gradient in Fig. \ref{fig:scatter-f-b0} indicates that the terminal wind speed is anti-correlated with $f_{tot}$; the transition from fast to slow wind is almost horizontal (or orthogonal to the $f_{tot}$ axis).
But there also are clear variations of wind speed in the direction of the $B_0$ axis (which are particularly visible in the slow wind part of the diagram, plotted in different shades of blue).
The figure therefore indicates that these two parameters should indeed be combined in order to predict the solar wind speed.

Figure \ref{fig:scatter-v-b0f} relates the terminal wind speed $V_{wind}$ with the ratio $B_0/f_{tot}$ for all the data points in our simulations.
The figure shows three curves which represent the power-law $V_{wind} \propto \left( B_0/f_{tot} \right)^{\nu}$ with three distinct indices $\nu$.
The green dashed line corresponds to $\nu = 0.5$, as proposed by \citet{suzuki_forecasting_2006}, the brown dashed line to $\nu = 0.1$, and the continuous blue line to $\nu=2.2$.
The blue and orange lines are empirical fits to the upper (fast wind) and lower (slow wind) envelopes of our data points.
All three curves are normalised and offset to arbitrary coefficients. 
It is clear from the figure that we cannot fit a single power-law of the kind $V_{wind} \propto \left( B_0/f_{tot} \right)^{\nu}$ to our data set.
We could perhaps fit a piece-wise power-law with at least two indices, one for the slow wind and the other for the fast wind.
However, it would still be necessary to specify where the transition between both should be placed as a function of the activity cycle and/or of the level of complexity of the background coronal field.
A small subset of our wind solution (moderately fast winds) seems to follow a curve corresponding to the intermediate index $\nu=0.5$, even though requiring different normalisation and offset coefficients at each given moment of the cycle.
This suggests that additional parameters should be taken into consideration.

\subsection{Influence of the field-line curvature}

\begin{figure}[!t]
  \centering
  \includegraphics[width=0.80\picwd,clip,trim=0 60 0 0]{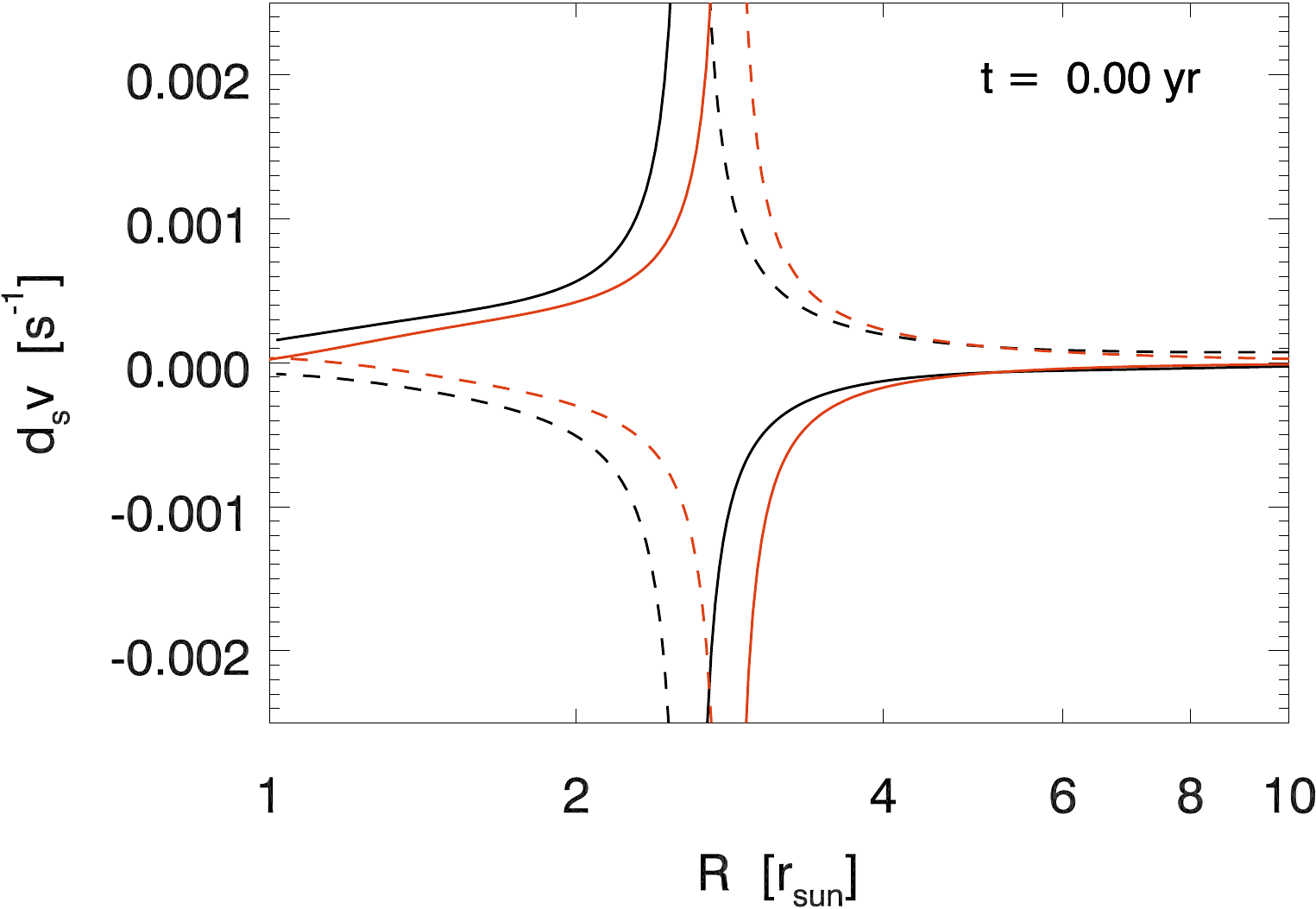}
  \includegraphics[width=0.80\picwd,clip,trim=0 60 0 0]{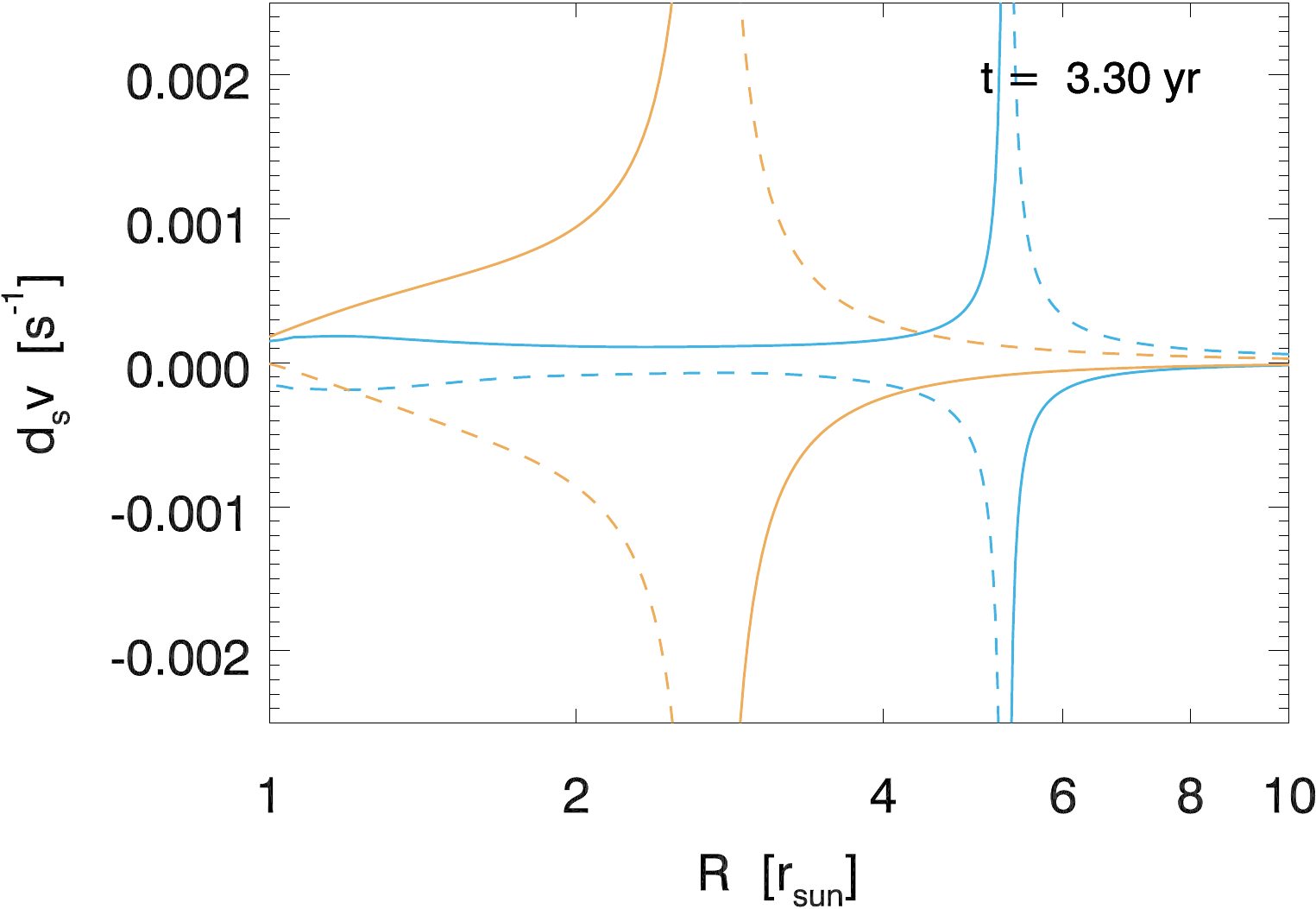}
  \includegraphics[width=0.80\picwd,clip,trim=0  0 0 0]{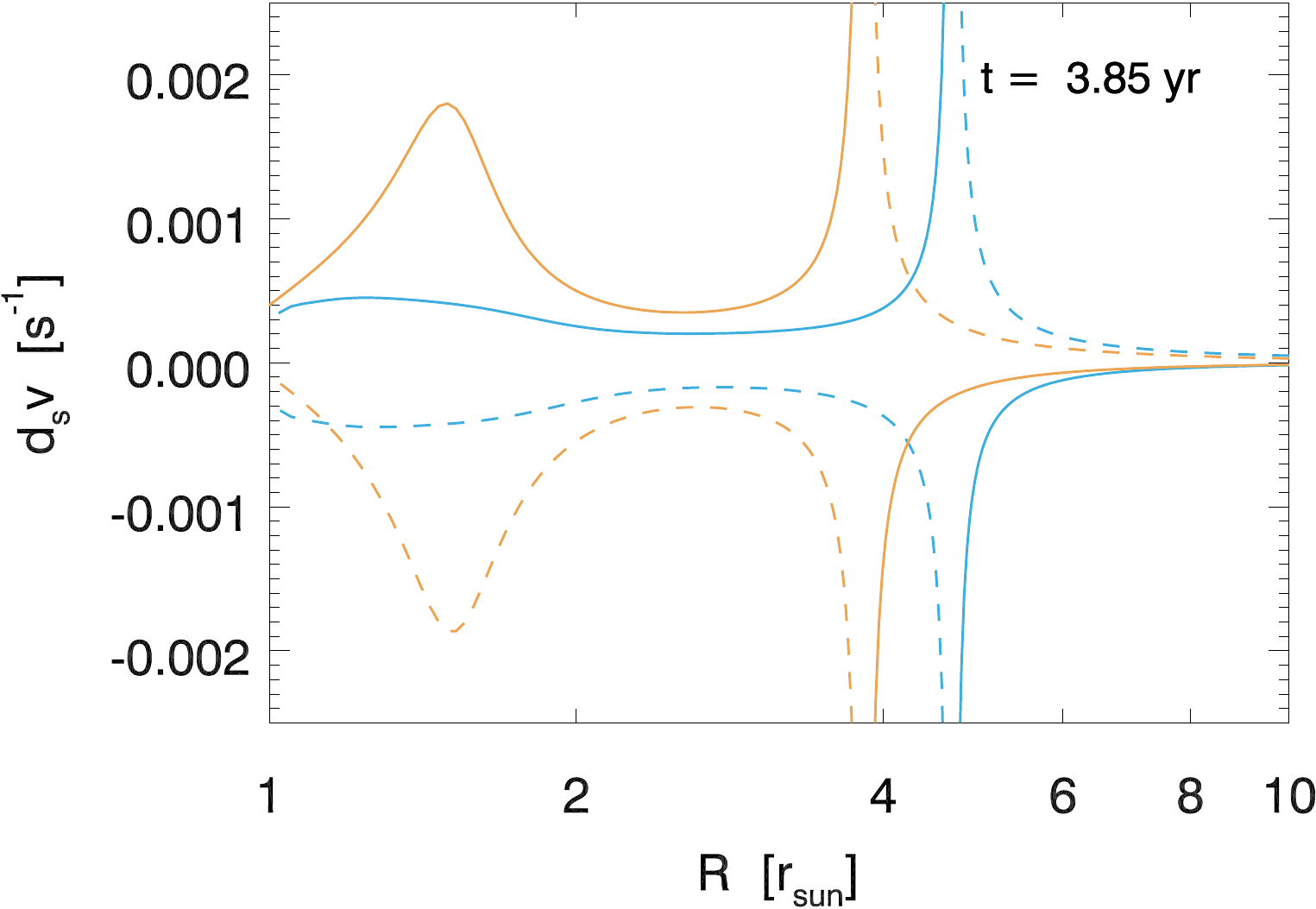}
  \caption{
    The two terms in the r.h.s. of Eq. (\ref{eq:mom_geom}) as a function of the distance to the surface for the flux-tubes which carry the fastest and the slowest wind flow at three different instants of the cycles (from the top to the bottom panel: $t=0,\ 3.30,\ 3.85\un{yr}$), as in Fig. \ref{fig:profiles-duterm-ratio}.
    The colour scheme (for the wind speed) is the same as in Figs. \ref{fig:expans-distance} and \ref{fig:profiles-inclin}.
    The continuous lines represent the first term (gravitation and inclination), and the dashed lines represent the second term (expansion).
    The terms switch sign at the sonic point (\emph{cf.} the denominator in Eq. \ref{eq:mom_geom}) at a height varying between $2$ and $6\un{\rsun}$.
  }
  \label{fig:profiles-duterms}
\end{figure}
\begin{figure}[!t]
  \centering
  \includegraphics[width=0.75\picwd,clip,trim=0 60 0 0]{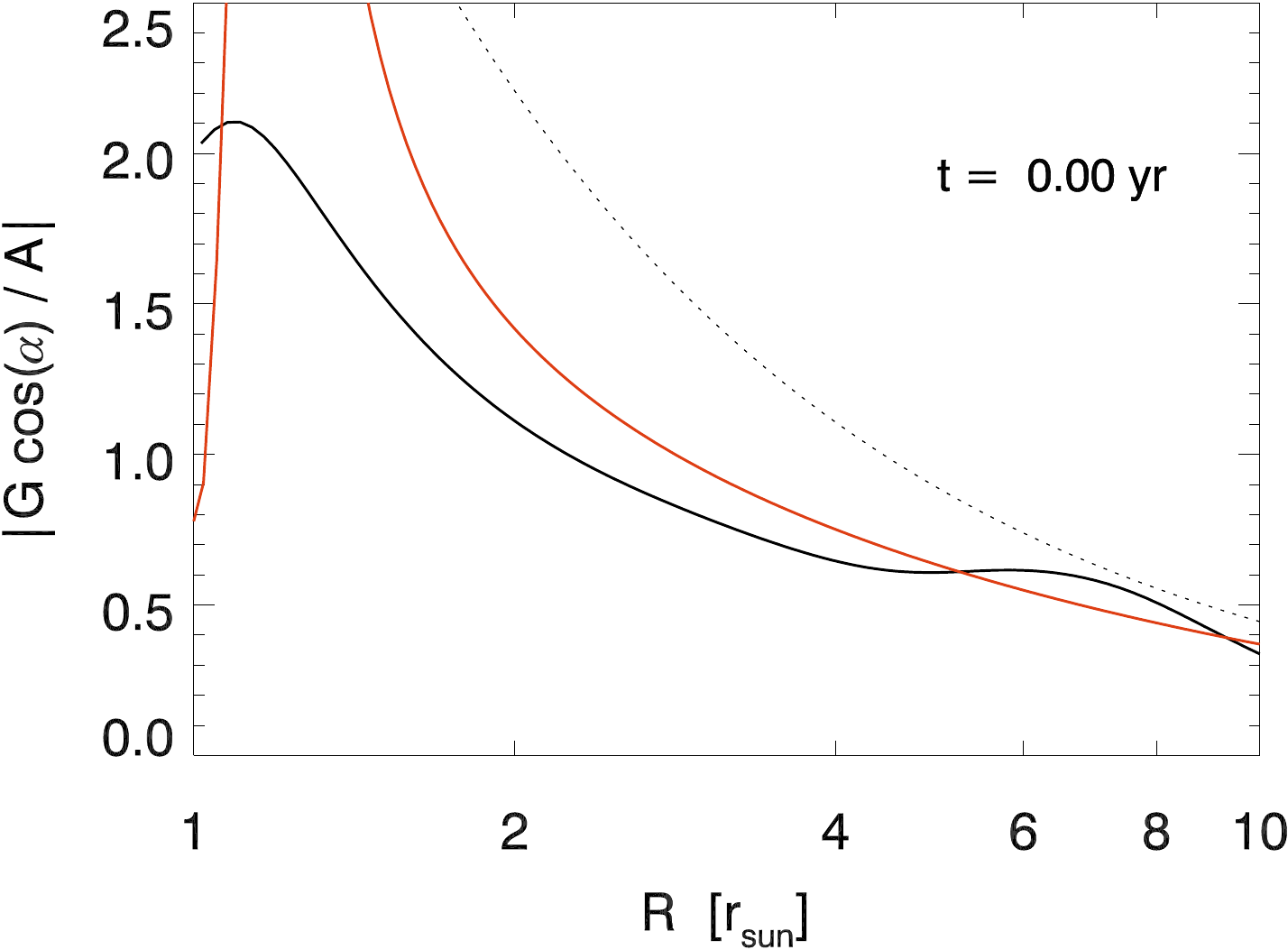}
  \includegraphics[width=0.75\picwd,clip,trim=0 60 0 0]{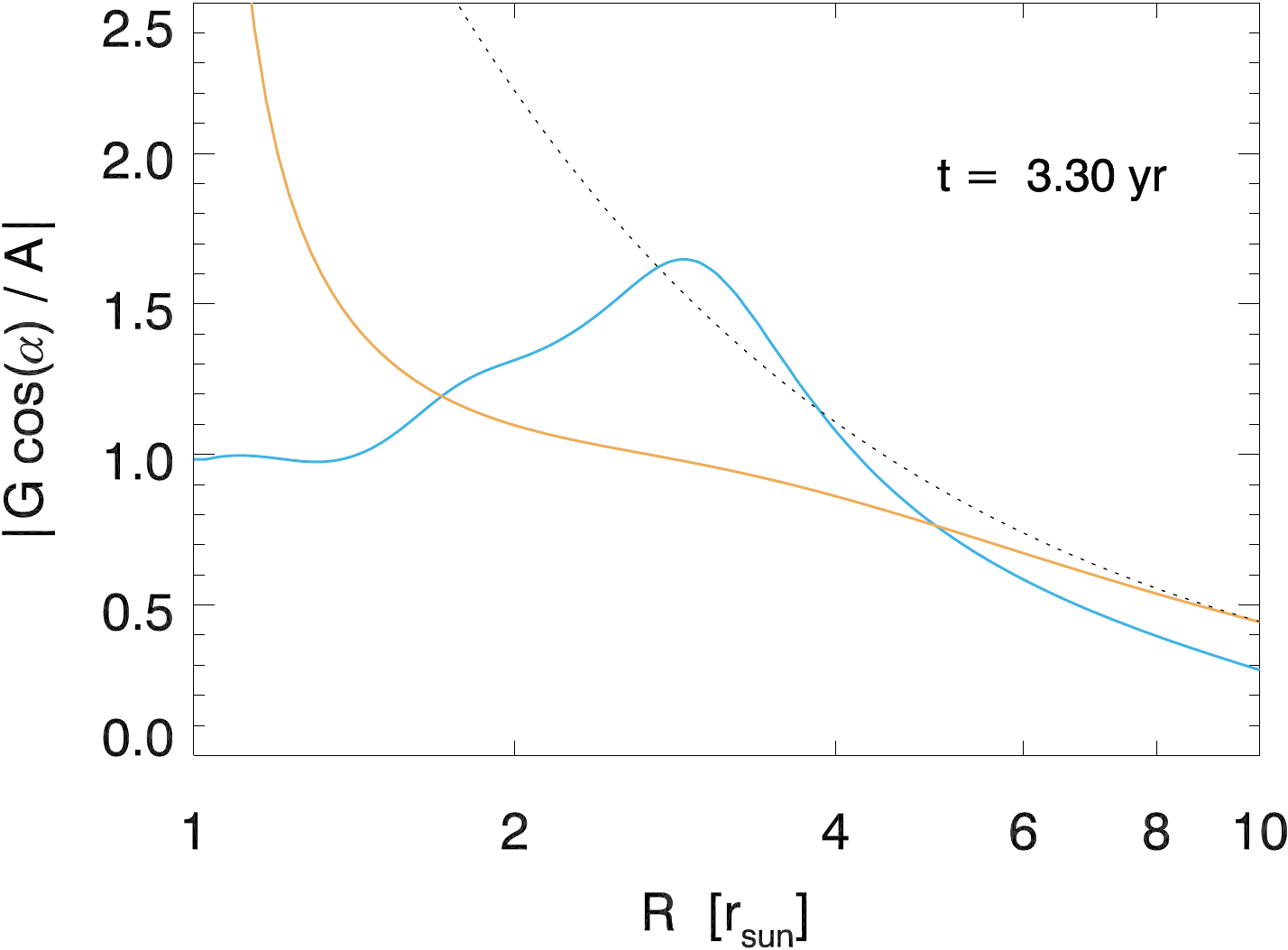}
  \includegraphics[width=0.75\picwd,clip,trim=0  0 0 0]{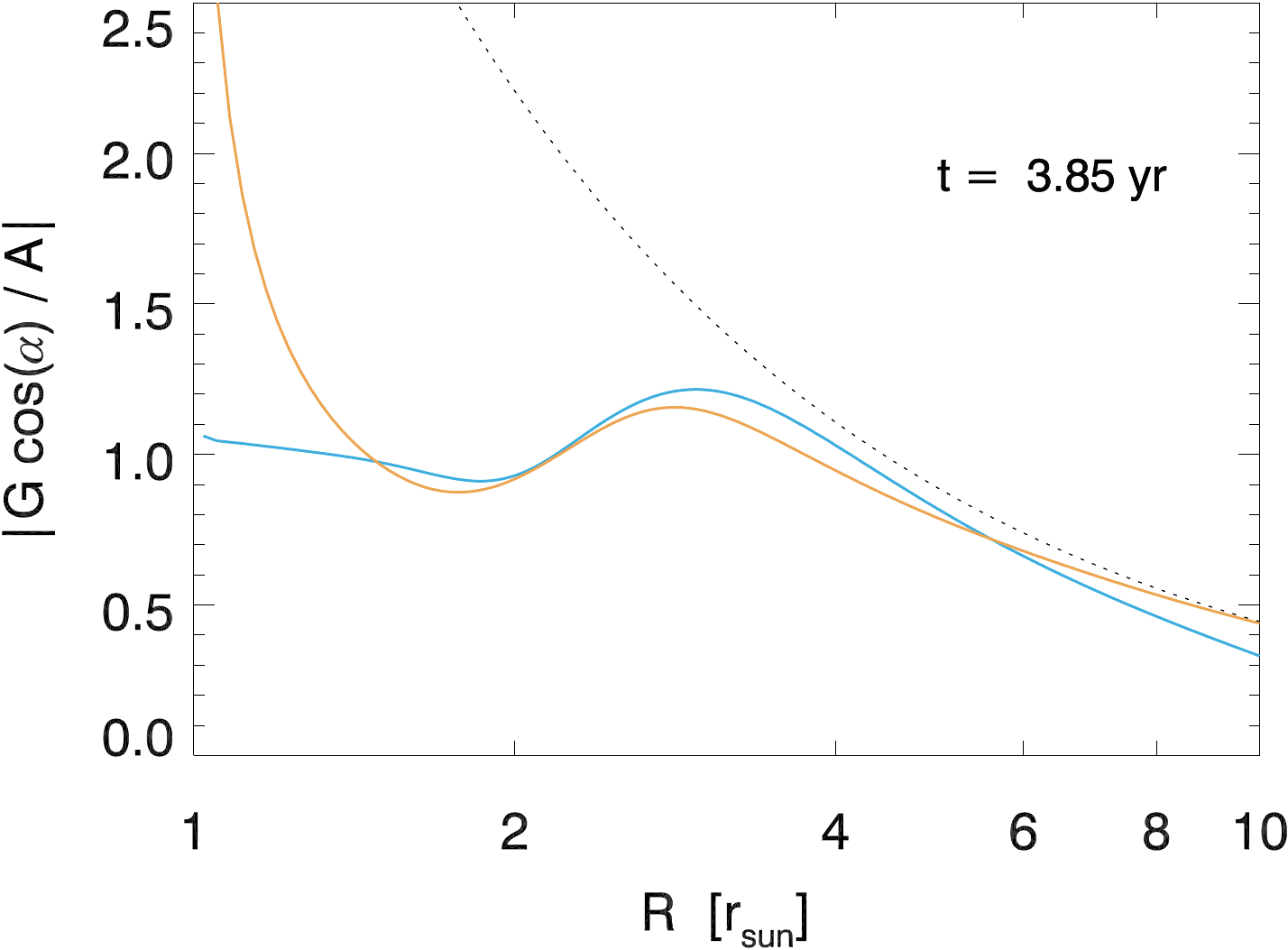}
  \caption{
    Ratio of the absolute value of the two terms in the r.h.s. of Eq. (\ref{eq:mom_geom}) as a function of the distance to the surface for the flux-tubes which carry the fastest and the slowest wind flow at three different instants of the cycles (the same as in Fig. \ref{fig:profiles-duterms}).
    The black dotted line shows the value of this ratio for a vertically-aligned and radially expanding wind flow (a Parker wind) for comparison.
  }
  \label{fig:profiles-duterm-ratio}
\end{figure}

We will now investigate the influence of the field-line curvature and inclination on the wind speed.
Large deviations of the flux-tubes from the vertical direction are expected to have several effects on the wind flow. 
The wind flowing along inclined portions of a flux-tube will see a reduced effective gravity (which can in extreme cases be null, or even flip sign on field-line switch-backs).
Field-line curvature will also increase the flow path length between two given heights (\emph{e.g} between the surface of the Sun and $1\un{AU}$) and decrease the pressure gradient felt by the wind flow.

Figure \ref{fig:profiles-inclin} shows the inclination of a large sample of field-lines as a functions of radius.
The inclination angle is the angle between the vertical direction and the direction parallel to the magnetic field at any given point of a field-line.
The different panels represent different instants of the cycle and orange/blue lines represent fast/slow winds, as in Fig. \ref{fig:expans-distance}.
The diamonds indicate the positions of the maximal inclination of each field-line represented.
The field-line inclination profiles vary considerably throughout the activity cycle.
The low latitude field-lines are significantly inclined in respect to the vertical direction from the surface up to more than $8\un{\rsun}$ during the minimum, but only up to $3 - 4 \un{\rsun}$ during the maximum.
This difference relates to the height of the streamers at those moments (see Fig. \ref{fig:intro-model-snaps}).
Field-lines passing close to the pseudo-streamer boundaries display the most extreme inclinations (reaching $\pi/2$ in some cases), as seen in the third and fifth panels of Fig. \ref{fig:profiles-inclin} (the third and fifth panels of Fig. \ref{fig:intro-model-snaps} show the geometry of these structures, respectively mid-latitude and polar pseudo-streamers).
In all cases, field-lines with large inclinations are systematically related to slower wind speeds than field-lines with small deviations to the vertical direction.
The only exceptions to this rule are field-lines with strongly inclined parts at lower coronal heights, below $\sim 1.5\un{\rsun}$.
In other words, the effect of field-line inclination seems to be important at coronal heights at which the wind flow has already started being accelerated -- reaching at least $25$ to $50\%$ of its terminal speed -- rather than at heights at which the wind speed is still negligible.
Flux-tubes which have longer inclined portions always have the slowest wind flows (the darker blue lines in Fig. \ref{fig:profiles-inclin}).

Figure \ref{fig:hist_i_v} shows three histograms of the terminal wind speed for three contiguous intervals of the maximum inclination angle $\alpha$ for each of the flux-tubes.
The black bars correspond to the interval $0^\circ \leq \alpha < 15^\circ$, the blue bars to $15^\circ \leq \alpha \geq 30^\circ$, and the red bars to $\alpha \geq 30^\circ$.
The figure shows that the slowest wind flows are always associated with flux-tubes with large deviations from the vertical direction, and that field-line inclination is generally anti-correlated with terminal wind speed.
However, as in Fig. \ref{fig:hist_f_v}, the spreads in the distributions are large and the different distributions overlap each other.
It must be noted, however, that extreme inclinations (\emph{e.g}, $\alpha \geq 60^\circ$) can be associated with moderate and fast wind flows, as long as the field-lines are bent only very low down in the corona (see Fig. \ref{fig:profiles-inclin}), where the wind speeds are still very small and effect of inclination on the wind speed is minimised.

Figure \ref{fig:speed-inclin} shows the terminal wind speed $v_{wind}$ plotted against $\left(B_0/f_{tot}\right)^{0.1}\times L \cos\left(\alpha\right)$, where the ratio $B_0/f_{tot}$ is the same as in Fig. \ref{fig:scatter-v-b0f}, $L$ is the total field-line length from the surface up to $r=15\un{\rsun}$, and $\alpha$ is the maximum inclination angle attained by the flux-tube.
This empirically derived expression reduces the scatter along the x-axis in the diagram in Fig. \ref{fig:scatter-v-b0f}, especially for the fast wind part of the diagram.
The slow wind part still maintains a large dispersion, nevertheless.

\section{Discussion and conclusions}
\label{sec:discussion}



This study confirms that the terminal (asymptotic) speed of the solar wind depends very strongly on the geometry of the open magnetic flux-tubes through which it flows.
Our results indicate that the flux-tube inclination should be added to the well-know and often invoked flux-tube expansion factor on future predictive laws for the solar wind speed. 
In our simulations, the wind speed is mildly anti-correlated with the total expansion factor (denoted $f_{tot}$ here, to be distinguished from the $f_{ss}$ used on observational studies based on PFSS extrapolations), but in a way which does not allow a general fit to a power-law of the kind $V_{wind} = a + b\cdot f^{\alpha}$ (see Fig. \ref{fig:f_vs_v}).
Furthermore, the spread in this relation is large, and a given expansion factor can correspond to different values of wind velocity, especially close to solar maximum (see Figs. \ref{fig:hist_f_v} and \ref{fig:expans_lat}).
While the variations in speed in the fast wind regime (especially during solar minimum) seem to be effectively controlled by the flux-tube expansion, those in the mid to slow wind regimes require the additional deceleration provided by the field-line inclination in the low to mid corona.

To get an insight on the combined effect of these two geometrical parameters (flux-tube expansion and inclination), let us express the momentum equation of a wind flow along the direction of an arbitrary field-line as
\begin{equation}
  \label{eq:mom_generic}
  \rho v \partial_l v = -\rho g\cos{\left(\alpha\right)} - \partial_l P + F\ ,
\end{equation}
where $l$ is the coordinate parallel to the magnetic field, $\rho$ is the plasma density, $v$ the wind speed, $g = GM_{\odot} / r^2$ is the gravitational acceleration vector, $\alpha$ is the angle between the magnetic field and the vertical direction, $P$ is the gas pressure and $F$ the combination of any other source terms involved.
The inclination of the flux-tube intervenes directly on Eq. (\ref{eq:mom_generic}) by reducing the amplitude of the gravitational term, while the expansion acts by modulating all the fluxes at play (such as the mass flux, involving the density $\rho$ and the wind speed $v$).
The relative influence of the expansion and inclination of the flux-tube on the wind acceleration and deceleration becomes more clear if Eq. (\ref{eq:mom_generic}) is rewritten as
\begin{equation}
  \label{eq:mom_geom}
  \partial_l v = \frac{v}{1 - M^2} \left[ \frac{G M_{\sun} m_p}{2 k_b T r^2}\cos{\left(\alpha\right)} - \frac{1}{A}\partial_l A \right]\ ,
\end{equation}
where $M$ is the wind Mach number, $T=T_0$ is the plasma temperature, $A$ is the flux-tube cross-section, and the other symbols have their usual meanings \citep[\emph{cf.}][]{wang_two_1994}.
The wind is assumed to be isothermal and composed of an ideal and fully ionised hydrogen gas, and other source terms (summarised by the term $F$ in Eq. \ref{eq:mom_generic}) were discarded for simplicity.
Below the sonic point (where $M < 1$), moderate inclination angles ($0 \leq \alpha  < 90\degree$) contribute to decelerating the wind flow by reducing the amplitude of the gravitational term.
More extreme situations with angles $\alpha > 90\degree$ (e.g, in field-line switchbacks) would lead to a local acceleration of the flow (although the net contribution to the terminal wind speed should be negligible, because of the deceleration occurring in the turning back of the field-line into the upward direction).
A growing flux-tube cross-section -- with a positive $\partial_l A\left(l\right)$ -- will also contribute to decelerating the wind flow.
The wind flow will be less decelerated (or even accelerated) on a ``re-converging'' section of a flux-tube.
Overall, the amplitude of these effects will be minimised altogether on sections of a flux-tube where the wind is either very slow ($v \approx 0$) or very fast ($M^2 >> 1$), and will be maximised at wind speeds close to the sound speed.
The flux-tube inclination and expansion terms change sign in the supersonic regime, where $M > 1$, hence reversing their effects on the acceleration/deceleration of the wind flow there.

Figure \ref{fig:profiles-duterms} shows the contributions of the two terms on the r.h.s. of Eq. (\ref{eq:mom_geom}) for the wind acceleration and deceleration $\partial_l v$, for the flux-tubes with maximal and minimal terminal wind speed. 
We denote these terms by $G\cos\left(\alpha\right)$ and $A$ hereafter, for simplicity.
The continuous lines represent the term $G\cos\left(\alpha\right)$, and the dashed lines represent the term $A$.
The colours represent the terminal wind speed, as before.
As explained above, the effect of both terms is maximal in the vicinity of the sonic point (where $M=1$, and both terms switch sign), and minimal close to the surface and at greater coronal heights.
The dashed and continuous lines are nearly symmetrical at all times, indicating that one term cannot in general be neglected in respect to the other.
Figure \ref{fig:profiles-duterm-ratio} represents the ratio of the absolute value of the two terms at the same instants as in Fig. \ref{fig:profiles-duterms}.
For comparison, this ratio is equal to $\left(G M_{\sun} m_p\right)/\left(4 k_b T r \right) $ for a vertically-aligned and radially expanding flux-tube, which would drive a Parker wind (represented by a black dotted line in the figure). 
The blue and orange lines (our solutions) approach this limiting case at high coronal altitudes, where flux-tube inclination and over-expansion become negligible.
The two terms are indeed at all times of the same order of magnitude, with the $G\cos\left(\alpha\right)$ being usually larger than $A$ (but rarely by a factor larger than $2$).
The slowest wind streams are, in general, associated with rather irregular curves for the $G\cos\left(\alpha\right)$; the dips in these curves are due to local increases in field-line inclination (or decreases in $\cos\left(\alpha\right)$).
Figures \ref{fig:expans_lat} and \ref{fig:expans-distance} indeed show that the wind speed is generally well anti-correlated with the expansion ratios for the fast wind flows close to the minimum of activity, but not so much for the slow wind nor during the maximum.
Fig. \ref{fig:scatter-f-b0} shows that in slow wind regime (plotted in shades of blue) the wind speed can actually vary more strongly with the magnetic field amplitude $B_0$ than with $f_{tot}$.
Some of the peaks in Fig. \ref{fig:expans_lat} (especially on the fourth panel) furthermore show that exceptions to the $f_{tot}$ -- $V_{wind}$ relation exist even in the fast wind regime.
All these deviations, \emph{i.e} winds which are slower than expected for their total expansion ratios, are related to flux-tubes which have strong inclinations in range of heights $r = 2 - 4\un{\rsun}$ (\emph{cf.} Fig. \ref{fig:profiles-inclin}).
Inclination does not have a significant impact on the wind flows when it occurs below these heights, as the wind flow is still very subsonic there (see Eq. \ref{eq:mom_geom}).
The slowest of the wind flows found exist in flux-tubes for which the inclined portions are particularly long (note the dark blue lines in the first panel of Figure \ref{fig:profiles-inclin}).

These results motivate the search for predictive wind speed scaling laws which use a combination of the geometrical parameters discussed here: flux-tube expansion, magnetic field amplitude and inclination. 
The plot in Figure \ref{fig:speed-inclin} illustrates a first step in this direction, and consists of a factor similar to that suggested by \citet{suzuki_forecasting_2006} corrected by $L\cos\left(\alpha\right)$, where $\alpha$ is the maximum inclination angle for a given flux-tube (marked on Fig. \ref{fig:profiles-inclin} with diamonds), and $L$ its length (which is a rough measure of the height interval over which the field-line is bent).
The resulting expression provides a much better fit to our wind solutions, even though some scatter remains on the slow wind part.
It is clear that the relative heights of the inclined and over-expanding fractions of the flux-tubes in respect to the height of the sonic point (or rather, the factor $v/\left(1-M^2\right)$ in Eq. \ref{eq:mom_geom}) is a key parameter regulating the asymptotic wind speeds.
But, unfortunately, the positions of the sonic point is not know in advance (they depend on the wind speed profile) and cannot be used as a predictor.

The analysis carried out in this manuscript is based on results of a series of numerical simulations using an MHD model of an axisymmetric solar corona at fixed temperature $T_0$.
The choice of the parameter $T_0$ has a direct effect on the wind solutions: a higher coronal temperature will produce a globally faster and denser wind \citep{parker_dynamical_1964-2,leer_constraints_1979,hansteen_coronal_1995,pinto_coupling_2011}.
However, the model will still reproduce the relative variations of the properties of the wind in latitude correctly (as long as $T_0$ remains fixed throughout and set to a reasonable value, in the range $1 - 1.5 \un{MK}$).
Fast and slow solar wind streams are known to have different temperature profiles, with slow winds reaching higher temperatures in the low corona than fast winds, but lower temperatures faraway from the Sun. 
The effect of these temperature variations could perhaps be simulated in our model by setting a non-uniform coronal $T_0$ (varying in latitude and in time). 
But, as discussed before, the distribution of fast and slow wind flows is the outcome of the model, on which its parameters cannot depend.
An in-depth investigation of the thermodynamics of the solar wind flows requires a different kind of model \citep[as in][]{suzuki_why_2006,pinto_time-dependent_2009,schwadron_solar_2003,woolsey_turbulence-driven_2014}, which will be the subject of future work.
Notwithstanding, the approach we adopt here strongly suggests that the absolute values and spatial distribution of the solar wind speed is mostly controlled by the geometry of the coronal magnetic field (independently of the heating scenario considered).

The coronal magnetic field is obtained via a kinematic mean-field dynamo model, and is not meant to reproduce a specific solar cycle.
Rather, it produces many coronal features representative of the evolution of any solar cycle.
More importantly, our simulations provide a very large statistical ensemble of open magnetic flux-tubes with wind flows covering a very large range of flux-tube geometries.

\section{Summary}
\label{sec:summary}

We have investigated how the solar wind speed relates with flux-tube geometry at all latitudes and moments of the solar cycle by means of global-scale MHD simulations of the solar dynamo, corona and wind.
The model generates maps of the slowly varying coronal magnetic field and of the wind velocity in the meridional plane from the solar surface up to $15\un{\rsun}$ during an $11\un{yr}$ activity cycle.
We analysed a large sample of individual magnetic flux-tubes covering the full latitude and time intervals in order to derive correlations between their geometrical parameters and the resulting wind speeds that remain valid for all latitudes and moments of the cycle.
We found that, in addition to the total expansion factors $f_{tot}$ and absolute magnetic field amplitudes $B_0$, the wind speed also depends strongly on flux-tube inclination.
Future work will focus on improving the empirically-fitted scaling law relating the wind speed to these three flux-tube parameters, on testing more sophisticated heating scenarios and on extending the analysis to sets of magnetic flux-tubes derived from observations.

\begin{acknowledgements}
  R. F. P. acknowledges funding by the FP7 project \#606692 (HELCATS).
  The dynamo--wind coupling method was developed within the frame of the STARS2 project (www.stars2.eu).
  A. S. B. acknowledges funding by the ERC grant Solar Predict, INSU/PNST and the CNES Solar Orbiter grant.
  We are grateful to Y.-M. Wang and R. Grappin for enlightening discussions and comments on the manuscript.
\end{acknowledgements}

   \bibliographystyle{aa}
   \bibliography{/data/rpinto/BIBTEX/refs,/data/BIBTEX/refs}

\end{document}